\documentclass[
 prd,preprint,longbibliography,
 showpacs,showkeys,lengthcheck,
 nofootinbib,tightenlines,onecolumn,notitlepage,
 preprintnumbers,superscriptaddress
]{revtex4-2}
\usepackage{amsmath}
\usepackage[utf8]{inputenc}
\usepackage{newtxtext,newtxmath}
\usepackage[mathcal]{euscript}
\usepackage{orcidlink}
\numberwithin{equation}{section}
\usepackage{graphicx} 
\usepackage{subcaption} 
\usepackage{booktabs}

\usepackage{hyperref}
\hypersetup{colorlinks=true,citecolor=blue,linkcolor=blue,urlcolor=blue}
\usepackage{orcidlink}

\begin{document}
\title{Nuclear Shell Model on the Light Front }
\author{Zhenkai Fu \orcidlink{????-????-????-????}\,}  
\email{eric722@uw.edu}
\affiliation{Department of Physics, University of Washington, Seattle, WA 98195-1560, USA}
\author{Adam Freese \orcidlink{0000-0002-0688-4121}}
\email{afreese@jlab.org}
\affiliation{Center for Nuclear Femtography, Southeastern Universities Research Association, Newport News, Virginia, 23606, USA}
\affiliation{Theory Center, Jefferson Lab, Newport News, Virginia, 23606, USA}
\author{Gerald A. Miller \orcidlink{0000-0003-2443-3639}}
\email{miller@uw.edu}
\affiliation{Department of Physics, University of Washington, Seattle, WA 98195-1560, USA}
\date{\today}

\begin{abstract}
    Light-front descriptions of nuclear structure provide a relativistic framework that can potentially deepen our understanding of nuclear structure by revealing effects that are absent from traditional non-relativistic treatments. To utilize this framework, we start from the light-front harmonic oscillator to analytically calculate the shell structure and electromagnetic form factors. The relativistic shell model shows that the light-front result recovers the nonrelativistic oscillator limit while introducing shell splitting that depends on the longitudinal excitation quantum number. We derive the resulting effects on the shell model and use the Hellmann–Feynman theorem to identify the longitudinal confinement interaction as the origin of this splitting. We illustrate these results through applications to \(^{48}\mathrm{Ca}\) and \(^{208}\mathrm{Pb}\). In addition, we demonstrate a substantial difference between form factors calculated using non-relativistic and light-front treatments, a difference that further underscores the importance of including relativistic effects in the description of nuclear structure. Our findings demonstrate a tractable approach to relativistic nuclear calculations and show how extending nuclear models beyond the non-relativistic approximation can yield meaningful physical insight.
\end{abstract}

\preprint{JLAB-THY-26-5044}
\preprint{NT@UW-26-28}

\maketitle

\section{Introduction}

In non-relativistic quantum mechanics, one can separate the internal and barycentric motion of a two-body system into two Hilbert spaces, which enables the resulting internal eigenvalue problem to have a direct three-dimensional interpretation.  This operation cannot be easily done in relativistic systems because boosts and spatial translations are part of the full Poincar\'e algebra, and the separation of internal and barycentric degrees of freedom is largely constrained \cite{Dirac:1949cp,Newton:1949cq}.  In particular, in instant-form  coordinates, a time-independent three-dimensional wave function that depends only on internal variables has been shown to be impossible \cite{miller2025impossibility}.  Relativistic harmonic-oscillator models have   been studied in several different formulations as controlled examples of relativistic bound-state dynamics \cite{Feynman:1971wr,Leutwyler:1977vy}.  A formulation in which the internal dynamics of a composite system can be defined independently of its barycentric motion is thus especially useful for studying the relativistic modification of familiar non-relativistic bound-state descriptions.

Light-front dynamics has provided such a framework.  Since its formulation as one of Dirac's forms of relativistic dynamics \cite{Dirac:1949cp}, light-front quantization has been developed extensively in relativistic field theory and bound-state physics \cite{Kogut:1969xa,Kogut:1972di,Brodsky:1997de,Miller:2000kv}.  Its kinematic structure permits a valid separation between internal and barycentric motions and has various applications in relativistic few-body systems~\cite{Bakker:1979eg,Carbonell:1998rj,Blunden:1999gq,Blunden:1999hy,Vary:2009gt}.  Furthermore, in light-front holographic approaches, it has been demonstrated that confining bound-state dynamics can be formulated in terms of boost-invariant internal variables \cite{Brodsky:2006uqa,Brodsky:2014yha}.
Most directly relevant to this paper,
Freese, Lashbrook and Miller~\cite{Freese:2026ulx}
recently developed a three-dimensional boost-invariant formulation for relativistically moving constituents,
using the harmonic oscillator potential constructed by Li, Maris, Zhao and Vary~\cite{Li:2015zda}.
This formulation was expressed entirely in internal light-front variables.
The purpose of this paper is to show the applications of the relativistic harmonic-oscillator description to familiar aspects of nuclear structure and analyze how the resultant quantities are modified based on non-relativistic descriptions.


Historically, the nuclear shell model based on an independent-particle description of the nucleus provides one of the central organizing frameworks for understanding nuclear structure.  In particular, the shell model explains the enhanced stability observed at specific proton and neutron numbers in terms of closed single-particle shells, thereby defining the nuclear magic numbers \cite{Mayer1949,Haxel1949,Wong1998}. Despite the existence of more sophisticated many-body models, the shell-model picture remains valuable as a simple model that connects observed nuclear properties with the symmetry of the single-nucleon Hamiltonian and its eigenvalue spectrum.  The harmonic oscillator plays a particularly useful role in this independent-particle description because it captures the near-equilibrium behavior of many interactions \cite{Wong1998}. Therefore, replacing the non-relativistic harmonic-oscillator spectrum with the light-front spectrum can be a natural way to demonstrate how familiar non-relativistic shell organization is modified by relativistic dynamics.

Electromagnetic form factors provide a natural probe of relativistic corrections to harmonic oscillator eigenstates because they are directly connected to the spatial and momentum structure encoded in the bound-state wave function and possess a relativistically invariant definition \cite{Miller:2009sg,Freese:2023jcp}. We therefore calculate electromagnetic form factors using the eigenstates of the light-front harmonic oscillator and compare them with form factors calculated using the corresponding non-relativistic harmonic oscillator.  Together, the shell-model analysis of the spectrum and the form-factor analysis of the wave functions provide two complementary ways of examining how the relativistic light-front description modifies familiar nuclear-structure physics.

The paper is organized as follows.
Section~\ref{sec:harmonic} summarizes the light-front harmonic-oscillator formalism
developed in Ref.~\cite{Freese:2026ulx}
with a definition of particular internal variables and the corresponding analytical solutions
for the invariant-mass eigenstates and eigenvalues.
Section~\ref{sec:independent} applies the resulting spectrum to the nuclear shell model
and analyzes its shell structure and its non-relativistic limit.
Section~\ref{sec:energy} studies the special properties of the rest-frame energy spectrum,
including light-front splitting, resultant cluster structure,
and higher-order corrections in the invariant mass to energy conversion.
Section~\ref{sec:ff} calculates light-front electromagnetic form factors
and compares them with their non-relativistic counterparts.
We finally conclude in Sec.~\ref{sec:end}.
The Appendix discusses the determination of the interaction strength scale
used in the nuclear applications for specific isotopes and the
corrections associated with the approximations entering the conventional shell-model estimate.

\section{Harmonic Potential in the Light-front Formalism}
\label{sec:harmonic}

In the instant-coordinate relativistic formulation, the internal and barycentric variables of a composite-particle system are coupled in its Hamiltonian, which makes the separation of an independent internal wave function generally impossible~\cite{miller2025impossibility}. In the light-front formulation, this issue is resolved because the transverse Galilean subgroup allows the invariant mass operator $M^2$ to be written as a purely internal eigenvalue problem.

In this section, the light-front coordinates and the corresponding internal variables for a two-body system are introduced.
Following the formalism of Freese, Lashbrook, and Miller (FLM)~\cite{Freese:2026ulx},
the light-front invariant mass operator $M^2$ with a harmonic oscillator interaction potential
from Li, Maris, Zhao and Vary (LMZV)~\cite{Li:2015zda}
and its solution are shown in the holographic variable.
The corresponding eigenfunctions and invariant-mass spectrum provide the theoretical basis for the analysis developed in the following sections.
\subsection{Light-front coordinates}
Light-front coordinates are defined by reparametrizing time and longitudinal displacement; the new spacetime four-vector is defined as:
\begin{equation}
x^\pm=\frac{1}{\sqrt{2}}(t\pm z),
\qquad
\mathbf{x}_\perp=(x,y).
\label{eq:lf_coordinates}
\end{equation}
We treat $x^+$ as the light-front time, and the momentum four-vector is defined as the generator of spacetime translations. Therefore, $p^-$, as the generator of $x^+$, acts like the new Hamiltonian. Starting from the relativistic shell condition, the Hamiltonian of one free particle can be written as:
\begin{equation}
m^2=2p^+p^- -\mathbf{p}_\perp^2,
\qquad
p^-=\frac{m^2+\mathbf{p}_\perp^2}{2p^+}.
\label{eq:lf_mass_shell}
\end{equation}

\subsection{Separation of variables}
One can use the Poincar\'e algebra in the light-front formalism to prove that a Galilei subgroup exists, which enables the algebra in the transverse direction to be analogous to 2D non-relativistic dynamics. In this Galilei subgroup, the $p^+$ operator acts like a central charge, which is analogous to mass in non-relativistic dynamics~\cite{Freese:2026ulx}. We treat the system as an effective two-body problem with constituent masses $m_1$ and $m_2$, where $m_1$ denotes the mass of the active constituent,
while $m_2$ denotes the mass of the remaining spectator system.

For a two-body system with constituent momenta $p_1$ and $p_2$, we define the barycentric variables as:
\begin{equation}
P^+=p_1^++p_2^+,
\qquad
\mathbf{P}_\perp=\mathbf{p}_{1\perp}+\mathbf{p}_{2\perp},
\label{eq:momentum_fraction}
\end{equation}
and the longitudinal and transverse internal variables are defined as:
\begin{equation}
    \qquad
x=\frac{p_1^+}{P^+},
\qquad
1-x=\frac{p_2^+}{P^+}
\end{equation}

\begin{equation}
\mathbf r_\perp
=
\mathbf x_{1\perp}-\mathbf x_{2\perp},
\qquad
\mathbf k_\perp
=
\frac{
p_2^+\mathbf p_{1\perp}
-
p_1^+\mathbf p_{2\perp}
}{
p_1^++p_2^+.
}
\label{eq:relative_variables}
\end{equation}
Then the invariant mass operator $M^2$ can be naturally expressed as follows:
\begin{equation}
M^2
=
2P^+P^--\mathbf P_\perp^2
=
M_{\mathrm{free}}^2+M_{\mathrm{int}}^2,
\label{eq:mass_operator}
\end{equation}
with
\begin{equation}
M_{\mathrm{free}}^2
=
\frac{m_1^2}{x}
+
\frac{m_2^2}{1-x}
+
\frac{\mathbf k_\perp^2}{x(1-x)}.
\label{eq:free_mass_operator}
\end{equation}

\subsection{Define $M_{\mathrm{int}}^2 $ for the harmonic potential}
Since $M^2$ is relativistically invariant and our goal is to separate the internal and barycentric variables, $M_{\mathrm{int}}^2 $ should only involve operators that are boost-invariant and independent of the barycentric position and momentum \cite{Freese:2026ulx}.
FLM~\cite{Freese:2026ulx} use the Miller–Brodsky variable~\cite{Miller:2019ysh} to describe longitudinal displacement: 
\begin{equation}
\widetilde z
=
P^+\left(x_2^- -x_1^-\right)
+
\mathbf P_\perp\cdot\mathbf r_\perp.
\label{eq:miller_brodsky}
\end{equation}
To solve the bound-state equation, we also introduce the Brodsky–de Téramond holographic transverse coordinate:
\begin{equation}
\zeta
=
\sqrt{x(1-x)}\,r_\perp.
\label{eq:holographic_coordinate}
\end{equation}
Taking into account hermiticity under the change in the integral measure that results from introducing the holographic variable, we define the holographic Miller–Brodsky operator by
\begin{equation}
\widetilde Z_\zeta
\psi(x,\zeta,\phi)
=
ix(1-x)
\frac{\partial}{\partial x}
\left[
\frac{\psi(x,\zeta,\phi)}
{x(1-x)}
\right]_{\zeta}.
\label{eq:holographic_MB_operator}
\end{equation}

Given all the above available displacement operators, we can define multiple light-front harmonic potentials, as the potential is not uniquely derived by separation of variables. Following FLM, we choose the LMZV interaction \cite{Li:2015zda} for the light-front harmonic oscillator:
\begin{equation}
M_{\mathrm{int}}^2
=
\kappa^4
\left[
x(1-x)r_\perp^2
+
\frac{1}{\Lambda^2}
\widetilde Z_\zeta
x(1-x)
\widetilde Z_\zeta
\right].
\label{eq:lf_ho_potential}
\end{equation}

Here $\kappa $ sets the strength of the confining interaction, while $\Lambda$ sets the longitudinal mass scale. For systems that possess a nonrelativistic limit, the appropriate choice is
\begin{equation}
\Lambda=m_1+m_2.
\label{eq:Lambda_definition}
\end{equation}

The transverse confinement potential is motivated by light-front holography with the soft-wall AdS/QCD model~\cite{Brodsky:2014yha}. The longitudinal confinement potential is adopted from LMZV \cite{Li:2015zda}, motivated primarily by the expected power-law behavior of parton distributions. 

A complementary physical motivation for the LMZV construction
is that it reproduces the non-relativistic harmonic oscillator potential
in the non-relativistic limit.
This motivation was stated already by LMZV~\cite{Li:2015zda}.
A lengthier pedagogical explanation,
which starts from the non-relativistic harmonic oscillator
and extends it to the light front,
was provided in an introductory lecture series by Freese~\cite{AdamLecture},
which we shall describe here.

In the nonrelativistic rest frame, the interaction is described by
\begin{equation}
V_{\mathrm{NR}}
=
\frac{1}{2}\mu\omega^2r^2
=
\frac{1}{2}\mu\omega^2
\left[
r_\perp^2+(z_1-z_2)^2
\right],
\end{equation}
where $\mu=m_1m_2/(m_1+m_2)$ is the reduced mass.
Since the light-front Hamiltonian eigenvalue problem is formulated
in terms of the invariant mass squared $M^2$, the interaction
can be expressed, to leading nonrelativistic order, as
\begin{equation}
\begin{aligned}
V_{M^2}
&\simeq 2(m_1+m_2)V_{\mathrm{NR}}\\
&=m_1m_2\omega^2
\left[
r_\perp^2+
\frac{\tilde z^2}{(m_1+m_2)^2}
\right],
\end{aligned}
\end{equation}
where $\tilde z$ is the Miller Brodsky variable mentioned above in Eq.~\eqref{eq:miller_brodsky}, which reduces to $(m_1+m_2)(z_1-z_2)$ in the
nonrelativistic rest frame.

When the interaction Hamiltonian is boosted to another frame, the rest-frame mass factor
$m_1m_2$ is generalized to $2(P^+)^2x(1-x)$. Meanwhile, motivated by
the red shift and blue shift of the oscillation frequency under longitudinal boosts, the squared frequency is assigned a scaling proportional to $(P^+)^{-2}$. These considerations motivate
the substitutions
\begin{equation}
\begin{aligned}
m_1m_2
&\longrightarrow
2(P^+)^2x(1-x),\\
\omega^2
&\longrightarrow
\frac{(m_1+m_2)^2}{2(P^+)^2}\omega^2.
\end{aligned}
\end{equation}
The explicit dependence on the total longitudinal momentum
$P^+$ consequently cancels, leading to the proposed
boost-invariant harmonic oscillator interaction
\begin{equation}
M_{\mathrm{INT}}^2
=
\kappa^4x(1-x)
\left[
r_\perp^2+
\frac{\tilde z^2}{(m_1+m_2)^2}
\right],
\end{equation}
where
\begin{equation}
\kappa^4=(m_1+m_2)^2\omega^2.
\end{equation}

This construction provides an intuitive physical motivation for the appearance of $x(1-x)$ in the light-front confinement potential while recovering the conventional three-dimensional
harmonic oscillator in the non-relativistic limit. However, $\tilde z$ does not commute with $x(1-x)$, so there is an ambiguity in first quantization.
Therefore, the precise quantum-mechanical ordering of the longitudinal position operators is specified separately by the LMZV construction \cite{Li:2015zda}.

\subsection{Light-front harmonic potential eigenstates and eigenvalue spectrum}

Using the holographic coordinate defined above, we obtain the final two-body holographic Schrödinger equation:
\begin{equation}
\begin{aligned}
x(1-x)
\Bigg[
&-\frac{\partial^2}{\partial \zeta^2}
-\frac{1}{\zeta}\frac{\partial}{\partial \zeta}
-\frac{1}{\zeta^2}\frac{\partial^2}{\partial \phi^2}
+\frac{m_1^2}{x}
+\frac{m_2^2}{1-x}
-M^2
+\kappa^4\zeta^2
\\
&-\frac{\kappa^4}{(m_1+m_2)^2}
\frac{\partial}{\partial x}
\left(
x(1-x)\frac{\partial}{\partial x}
\right)
\Bigg]
\frac{\psi(x,\zeta,\phi)}{x(1-x)}
=0 .
\end{aligned}
\label{eq:Schrodinger_equation}
\end{equation}
FLM solve this Schrödinger equation:
\begin{equation}
\psi(x,\zeta,\phi)
=
x(1-x)
X(x)
Z(\zeta)
\frac{e^{im_l\phi}}{\sqrt{2\pi}}.
\label{eq:separation}
\end{equation}
The resultant normalized eigenstates may then be expressed as
\begin{equation}
\begin{aligned}
\psi_{n_z n_\perp m_l}(x,\zeta,\phi)
={}&
\mathcal N_{n_z n_\perp m_l}
x^{(\alpha+2)/2}
(1-x)^{(\beta+2)/2}
\zeta^{m_l}
\\
&\times
P_{n_z}^{(\alpha,\beta)}(1-2x)
L_{n_\perp}^{m_l}
\left(
\kappa^2\zeta^2
\right)
e^{-\kappa^2\zeta^2/2}
e^{im_l\phi}.
\end{aligned}
\label{eq:lf_wavefunction}
\end{equation}
where
\begin{equation}
\alpha=\frac{2\Lambda m_1}{\kappa^2},
\qquad
\beta=\frac{2\Lambda m_2}{\kappa^2},
\label{eq:alpha_beta}
\end{equation}
and $P_{n_z}^{\alpha,\beta}$ are Jacobi polynomials. The corresponding invariant-mass eigenvalues are
\begin{equation}
M_{n_z,n_\perp,m_l}^2
=
\kappa^2
\left\{
4n_\perp
+
2m_l
+
2
+
\frac{\kappa^2}{\Lambda^2}
\left[
n_z^2
+
(\alpha+\beta+1)n_z
+
\frac{(\alpha+\beta)(\alpha+\beta+2)}{4}
\right]
\right\}.
\label{eq:lf_spectrum}
\end{equation}

Eq.~\eqref{eq:lf_spectrum} constitutes the principal result required for the spectral analysis developed in the following section. In contrast to the nonrelativistic isotropic harmonic oscillator, the exact light-front spectrum depends separately on the longitudinal and transverse quantum numbers. In particular, the quadratic dependence on $n_z$ generates a relativistic splitting of states that become degenerate in the nonrelativistic limit.

\begin{figure*}[t]
    \centering

    \begin{subfigure}{0.49\textwidth}
        \centering
        \includegraphics[width=\linewidth]{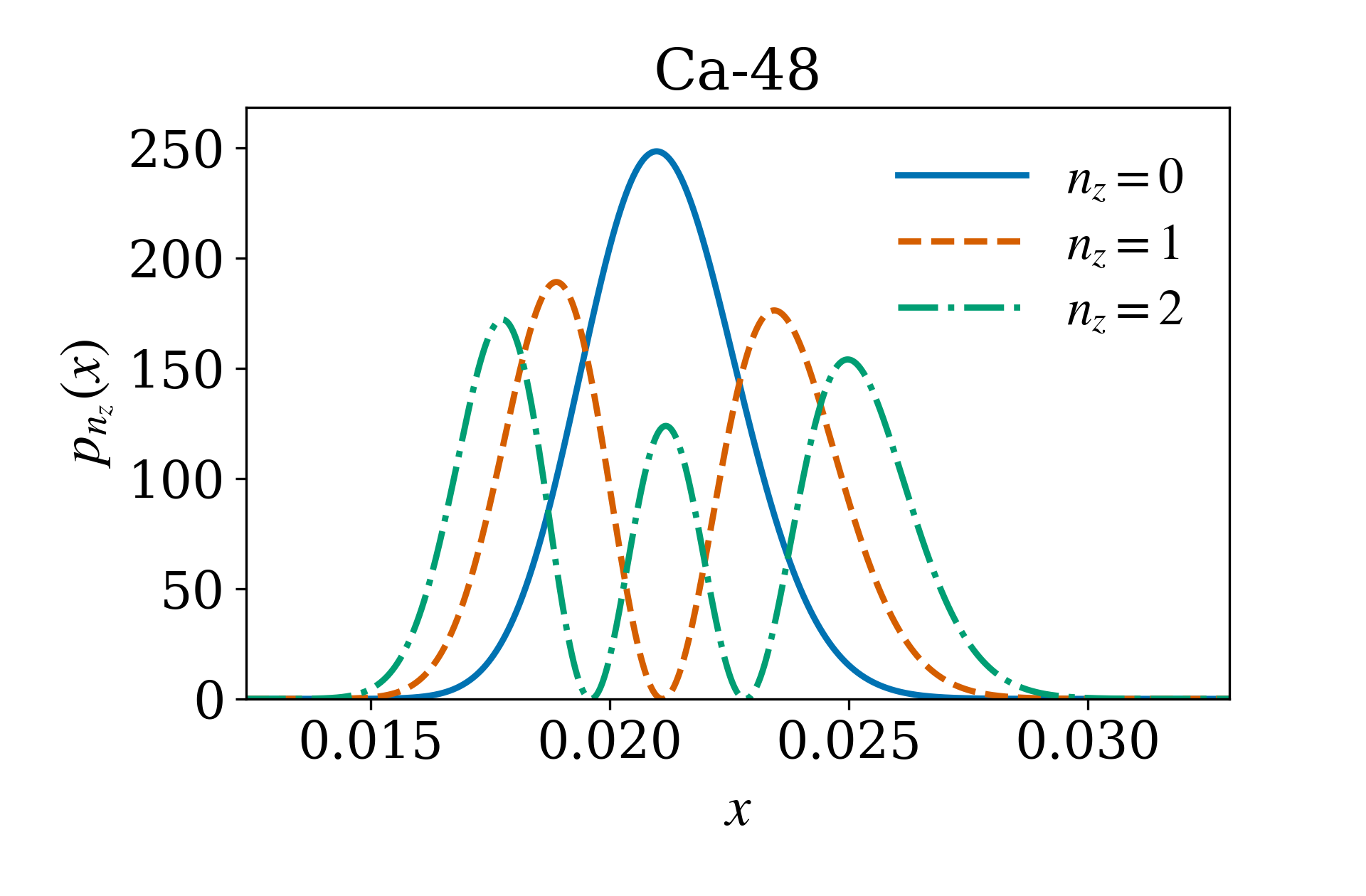}
        \caption{}
        \label{fig:a}
    \end{subfigure}
    \hfill
    \begin{subfigure}{0.49\textwidth}
        \centering
        \includegraphics[width=\linewidth]{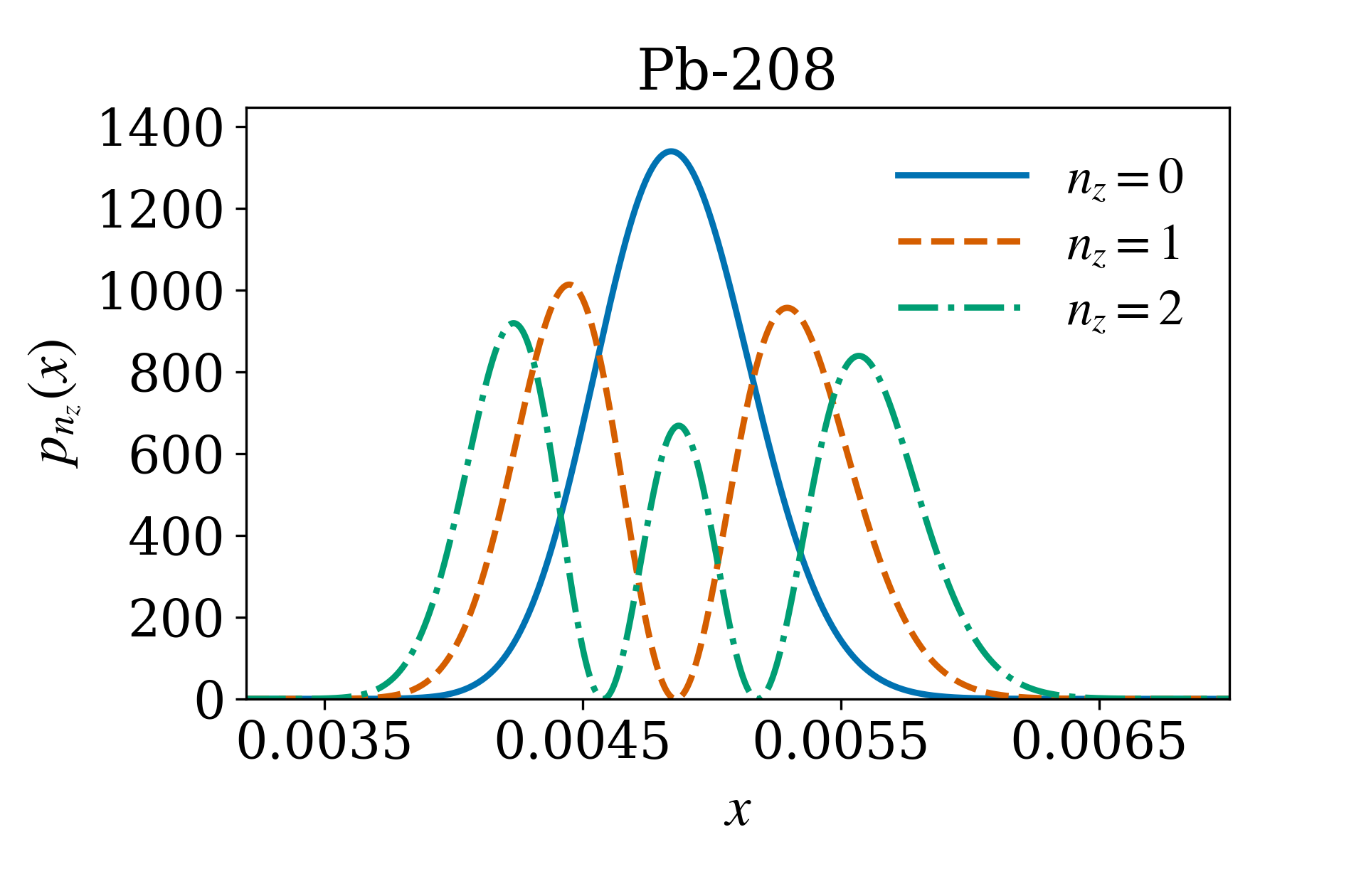}
        \caption{}
        \label{fig:b}
    \end{subfigure}

    \caption{Longitudinal momentum-fraction probability distributions $p_{n_z}(x)$ for the light-front harmonic-oscillator states with $n_z=0,1,$ and $2$. Results are shown for $^{48}\mathrm{Ca}$ in(a) and $^{208}\mathrm{Pb}$ in (b). The ground longitudinal mode, $n_z=0$, is concentrated around the most probable momentum fraction where momentum is distributed based on the mass ratio of the constituents, while increasing $n_z$ introduces additional nodes and a broader multi-peaked structure in the longitudinal momentum distribution.}
    \label{fig:comparison}
\end{figure*}

\section{Independent-particle model (nuclear shell model)}
\label{sec:independent}

For a nucleus as a quantum-mechanical many-body system made up of neutrons and protons, one natural way to simplify the analysis is the independent-particle model \cite{Wong1998}. Each nucleon is considered to move independently but to feel a two-body potential between it and the other nucleons in the nucleus. The harmonic potential is a natural interaction potential to choose, since it is a quadratic approximation of any interaction near the equilibrium points.

To apply the developed light-front harmonic oscillator potential in the independent-particle model, we consider the motion of a single proton
relative to the rest of the nucleus as an effective two-body system.  The
first constituent is therefore the proton, with mass
$m_1 = m_p $,
while the second constituent, with mass $m_2 \simeq (A-1)m_p$, is the residual $(A-1)$-nucleon system.
The corresponding light-front internal variables describe the relative motion between the proton and this spectator system. 
Because separation of relative and barycentric variables is possible on the light front,
the description of the internal state in this model
is independent of the location of the origin.

After choosing a two-body potential and neglecting residual interactions, we obtain the many-body nuclear Hamiltonian by summing the individual  single-nucleon terms. 
In the following analysis, the single-particle interaction is restricted to
the light-front harmonic potential. No explicit spin-orbit interaction is included. The spectrum analyzed below, therefore, isolates the modification
of the harmonic-oscillator shell structure produced by the relativistic light-front formulation itself. More work needs to be done to include the spin-orbit potential and corresponding energy splitting in the current model to obtain  a more complete description of the relativistic nuclear shell model.

\subsection{Degeneracy of the relativistic harmonic oscillator}

Closed-shell nuclei, which have numbers of nucleons that satisfy a group of ``magic numbers,''
share special features like enhanced stability. In the non-relativistic shell model, the observed sequence of magic numbers is
reproduced only after the spin-orbit interaction is included~\cite{Mayer1949,Haxel1949,Wong1998}.  Since no explicit spin-orbit
interaction is included in the present model, we first examine the shell structure generated by the relativistic
harmonic spectrum alone. Non-relativistically, the appearance of this shell closure is explained by the independent-particle model using the discrete nature of the single-particle energy spectrum given by the harmonic potential, and magic numbers can be regenerated by counting the degeneracy of each energy level after including the spin-orbit interaction. Therefore, examining the degeneracy of the light-front $M^2$ spectrum as a first step can provide valuable insight into the light-front independent-particle model.

The relativistic harmonic-oscillator mass-squared spectrum is given by Eq.~\eqref{eq:lf_spectrum}, which depends on the quantum numbers $n_z$ , $n_\perp$, and $m_l$ , which are non-negative integers. $m_\ell$ is the azimuthal quantum number; a state with $m_\ell>0$ has a two-fold degeneracy.

Since we treat the longitudinal coordinate separately, rotations that mix longitudinal and transverse coordinates acquire dynamical generators, which means that the full SO(3) rotational symmetry of the non-relativistic system is not manifest.  This treatment breaks the degeneracy that exists in the non-relativistic harmonic potential. Consequently, $n_z$  should be treated separately from the transverse quantum numbers.

We define a total transverse quantum number as:

\begin{equation}
N_\perp
\equiv
2n_\perp+m_\ell.
\label{eq:Nperp_definition}
\end{equation}
For fixed $N_\perp$ and $n_z$, all pairs $(n_\perp,m_\ell)$ have the same eigenvalue, so we can denote the $M^2$ eigenvalue using only two quantum numbers:
\begin{equation}
M^2_{N_\perp,n_z}
=
\kappa^2
\left\{
2(N_\perp+1)
+
\frac{\kappa^2}{\lambda^2}
\left[
n_z^2
+
(\alpha+\beta+1)n_z
+
\frac{1}{4}
(\alpha+\beta)(\alpha+\beta+2)
\right]
\right\}.
\label{eq:relativistic_HO_Nperp}
\end{equation}
The degeneracy is  obtained by counting the number of $(n_\perp,m_\ell)$ pairs available for a given $N_\perp$.

\subsubsection{Relativistic degeneracy calculation}

To denote the degeneracy associated with each value of $m_\ell$ while
taking the $m_\ell=0$ case into account, we write
\begin{equation}
g_m(m_\ell)
=
2-\delta_{m_\ell 0}.
\label{eq:angular_degeneracy}
\end{equation}
The total degeneracy for a fixed $N_\perp$ is therefore
\begin{equation}
g_\perp(N_\perp)
=
\sum_{n_\perp=0}^{\infty}
\sum_{m_\ell=0}^{\infty}
\left(2-\delta_{m_\ell 0}\right)
\delta_{2n_\perp+m_\ell,N_\perp}.
\label{eq:gperp_double_sum}
\end{equation}
We simplify the sum using Eq.~\eqref{eq:Nperp_definition} and
the non-negativity constraint on the quantum numbers.
\begin{equation}
g_\perp(N_\perp)
=
\sum_{n_\perp=0}^{\left\lfloor N_\perp/2\right\rfloor}
\left[
2-\delta_{N_\perp-2n_\perp,0}
\right].
\label{eq:gperp_single_sum}
\end{equation}
We then consider the even and odd cases separately. For even
$N_\perp$,
\begin{equation}
N_\perp=2k.
\end{equation}
There are $k$ states with $m_\ell>0$, each contributing two states,
while the final pair has $m_\ell=0$ and contributes one state. Hence
\begin{equation}
g_\perp(2k)
=
2k+1
=
N_\perp+1.
\end{equation}
For an odd transverse shell,
\begin{equation}
N_\perp=2k+1.
\end{equation}
In this case all $k+1$ allowed values of $m_\ell$ are positive and
each contributes two states. Thus,
\begin{equation}
g_\perp(2k+1)
=
2(k+1)
=
N_\perp+1.
\end{equation}
Consequently, for any $N_\perp$,
\begin{equation}
g_\perp(N_\perp)=N_\perp+1.
\label{eq:transverse_degeneracy}
\end{equation}
Therefore, the orbital degeneracy of one
exact relativistic level is
\begin{equation}
g_{\mathrm{rel}}(N_\perp,n_z)
=
N_\perp+1.
\label{eq:relativistic_orbital_degeneracy}
\end{equation}
More generally, for a   spin $1/2$ nucleon, the spin multiplicity is
$2 $, giving
\begin{equation}
g_{\mathrm{rel}}^{(1/2)}(N_\perp,n_z)
=
2(N_\perp+1).
\label{eq:relativistic_spin_degeneracy}
\end{equation}
 The cumulative number of states for spin $\frac{1}{2}$, up to a
maximum transverse quantum number $N_{\perp,\max}$ at fixed $n_z$, is
\begin{equation}
D_{\mathrm{rel}}=\sum_{N_\perp=0}^{N_{\perp,\max}}
g_{\mathrm{rel}}^{(1/2)}
(N_{\perp};n_z)
=
(N_{\perp,\max}+1)(N_{\perp,\max}+2).
\label{eq:relativistic_cumulative_transverse_spin_half}
\end{equation}
Therefore, the light-front eigenvalue spectrum has a smaller degeneracy than the non-relativistic form, but exactly the same degeneracy and cumulative state number as the 2D harmonic oscillator.  This indicates that cylindrical symmetry in the transverse direction is well preserved.

Another observation is that the cumulative number of states up to a maximum transverse quantum number $N_{\perp,\max}$ for the light-front harmonic oscillator has exactly the same form as the degeneracy of one 3D non-relativistic energy level with quantum number n:
\begin{equation}
    g_\mathrm{NR}^{(1/2)}
=
2\sum_{\text{allowed }\ell}(2\ell+1)
=
2\sum_{k=1}^{n+1}k
=
(n+1)(n+2)
\end{equation}

\subsubsection{Nonrelativistic limit}

In the non-relativistic limit, define the relativistic spectrum:
\begin{equation}
M^2
\simeq
\lambda^2
+
\kappa^2
\left[
3+
2\left(
2n_\perp+n_z+m_\ell
\right)
\right],
\label{eq:NR_limit_M2}
\end{equation}
where the ordinary three-dimensional oscillator-shell quantum number can be directly related to $N_\perp$ and $n_z$.
\begin{equation}
n
\equiv
2n_\perp+n_z+m_\ell
=
N_\perp+n_z.
\label{eq:NR_total_shell}
\end{equation}
For a fixed $N_\perp$, the transverse degeneracy of this sector is
\begin{equation}
g_\perp(n-n_z)
=
n-n_z+1.
\end{equation}
The full spatial degeneracy of the
three-dimensional shell in the non-relativistic limit is therefore
\begin{equation}
g_n
=
\sum_{n_z=0}^{n}
g_\perp(n-n_z).
\label{eq:NR_degeneracy_sum_over_nz}
\end{equation}
The sum is given by  
\begin{equation}
g_n
=
\frac{(n+1)(n+2)}{2}.
\label{eq:NR_spatial_degeneracy}
\end{equation}
Accounting for the nucleon spin 
gives
\begin{equation}
g_\mathbf{NR~limit}^{(1/2)}
=
(n+1)(n+2).
\label{eq:NR_spin_half_degeneracy}
\end{equation}

Thus, the nonrelativistic limit of the light-front treatment preserves the same degeneracy as the three-dimensional nonrelativistic harmonic oscillator.

\section{Light-front energy spectrum correction}
\label{sec:energy}

The special treatment of longitudinal dynamics in the light-front formalism divides  the eigenvalues of $M^2$  into different levels based on the value of $n_z$. To maintain the shell-closure explanation without requiring the non-relativistic limit,   the energy splittings caused by different values of $n_z$ should be small enough to enable light-front energy levels with the same $n$ but different $n_z$ to form a sub-shell cluster.

In this section, we analyze how the relativistic energy correction changes the energy spectrum in the rest frame and how that changes the interpretation of the independent-particle model.

\subsection{Light-front rest-frame energy $E_{\mathrm{LF}}^{(1)}$}
\label{sec:rest}

Rewrite the analytical form of $M^2$ in Eq.~\eqref{eq:lf_spectrum} in terms of the non-relativistic shell index n defined by Eq.~\eqref{eq:NR_total_shell} and using the values of  $\alpha$ and $\beta$ leads to
  $M^2$:
\begin{equation}
\begin{aligned}
M^2
&=
\kappa^2\left\{2(n-n_z+1)+\frac{\kappa^2}{\Lambda^2}\left[n_z^2+
\left(\frac{2\Lambda^2}{\kappa^2}+1\right)n_z\right.\right.
 \left.\left.+\frac{1}{4}\left(\frac{2\Lambda^2}{\kappa^2}
\right)\left(\frac{2\Lambda^2}{\kappa^2}+2\right)\right]\right\}
\\[4pt]
&=
\Lambda^2
+\kappa^2(2n+3)
+\frac{\kappa^4}{\Lambda^2}(n_z^2+n_z).
\label{eq:LF_M^2}
\end{aligned}
\end{equation}
We can already see that a special pattern with three separate terms is obtained after simplification, where the first term corresponds to the rest-mass squared, the second term is proportional to the NR energy, and the third term, which depends only on $n_z$, reveals the special treatment in the longitudinal direction.

\setlength{\parskip}{10pt}
Consider expressing $M^2$ in terms of the rest-frame bonding energy and the total mass of the nuclear constituents,
\begin{equation}
M^2
=
(\Lambda+E_{\mathrm{INT}})^2
=
\Lambda^2
+2\Lambda E_{\mathrm{INT}}
+E_{\mathrm{INT}}^2.
\end{equation}
Assuming  that the total mass is much larger than $E_\mathrm{INT}$,  making a first-order approximation and ignoring the $E_{\mathrm{INT}}^2$ term yields:
 \begin{equation}E_{\mathrm{LF}}^{(1)} 
=
E_{\mathrm{INT}}
=
\frac{M^2-\Lambda^2}{2\Lambda}.
\label{eq:energy_first_order}
\end{equation}
so that 
\begin{equation}
\begin{aligned}
E_{\mathrm{LF}}^{(1)}=
\frac{\kappa^2}{2\Lambda}(2n+3)
+
\frac{\kappa^4}{2\Lambda^3}n_z(n_z+1).
\end{aligned}
\end{equation}

We connect the interaction strength $\kappa$ with the frequency of harmonic oscillation $\omega$ in the non-relativistic limit \cite{Freese:2026ulx}:
\begin{equation}
\kappa^2=\Lambda\omega.
\label{eq:kappa_omega}
\end{equation}
Using the definition of $\kappa^2$,
\begin{equation}
\begin{aligned}
E_{\mathrm{LF}}^{(1)}(n,n_z)
&=
E_{\mathrm{NR}}(n)
+
\frac{\omega^2}{2\Lambda}n_z(n_z+1).
\label{eq:LF_Energy}
\end{aligned}
\end{equation}
The light-front energy can be expressed as the non-relativistic energy at a given NR shell number $n$ plus an additional energy correction term that depends on $n_z$ and is of order $\kappa^2/2\Lambda^3$. 
The additional dependence on the longitudinal quantum number $n_z$ follows from the exact longitudinal eigenvalue problem, in which the confinement operator
$
-\frac{\kappa^4}{\Lambda^2}\frac{d}{dx}
\left[x(1-x)\frac{d}{dx}\right]$
acts together with the constituent mass terms $m_1^2/x+m_2^2/(1-x)$. The resulting eigenstate is a Jacobi polynomial, whereas the non-relativistic harmonic-oscillator eigenstate involves Laguerre polynomials. This correction term breaks the degeneracy between states with different longitudinal excitations within a fixed non-relativistic oscillator shell. Further study regarding this correction will be shown in
Sec.~\ref{sec:origin}.

For $n_z=0$, the light-front energy level has the same energy as the NR level with the corresponding n. Since $n_z$ not only denotes the z-direction excitation but also denotes the quantum excitation index of the quantum pattern by which $P^+$ is distributed between the two constituents, at the lowest-energy $P^+$ distribution, the light-front energy is equal to that of a 3D non-relativistic harmonic oscillator in which one direction remains only in the ground state. As $n_z$ increases, more features in the $P^+$ distribution will induce a energy correction.

\begin{figure}[htbp]
    \centering
    \includegraphics[width=\linewidth]{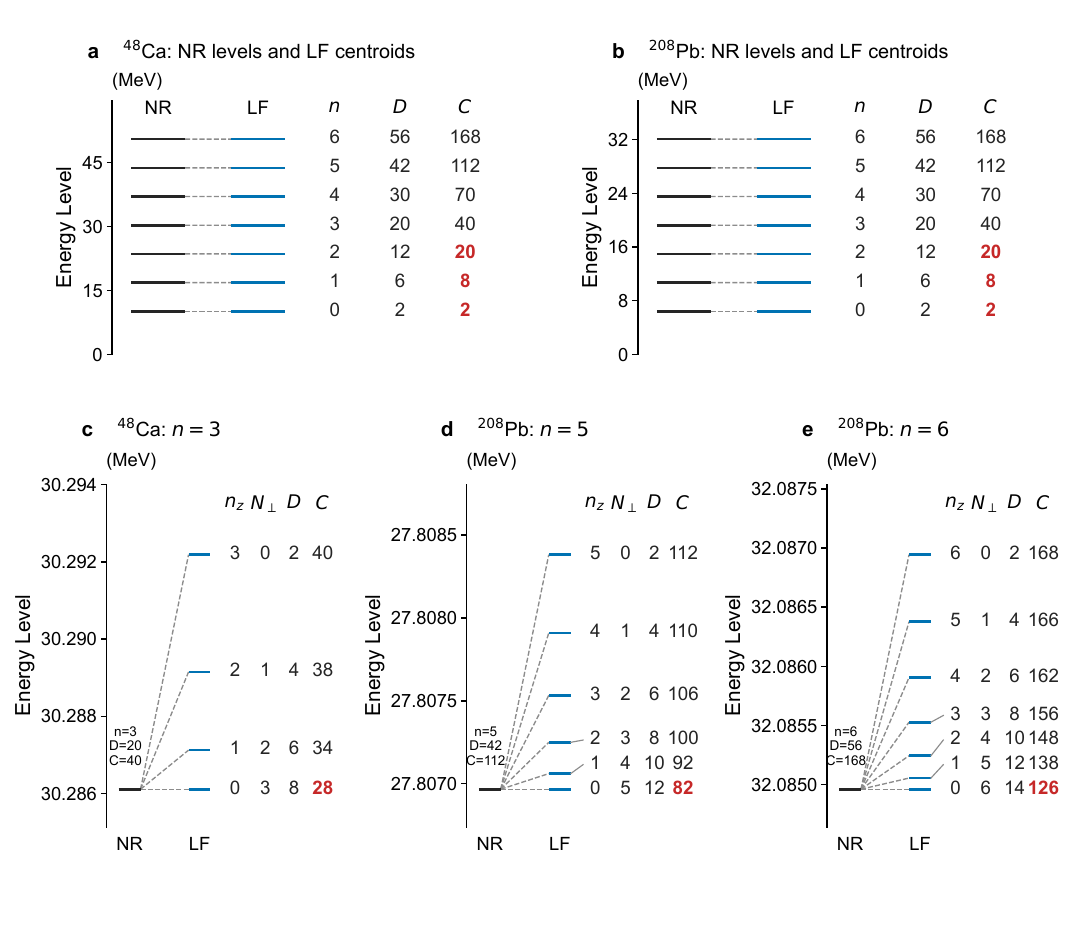}
    {\phantomsubcaption\label{fig:Ca48_centroid}}
    {\phantomsubcaption\label{fig:Pb208_centroid}}
    {\phantomsubcaption\label{fig:Ca48_n4}\label{fig:Ca48_n3}}
    {\phantomsubcaption\label{fig:Pb208_n5}}
    {\phantomsubcaption\label{fig:Pb208_n6}}

    \caption{
    Comparison of the nonrelativistic (NR) and light-front (LF)
    energy spectra for $^{48}\mathrm{Ca}$ and $^{208}\mathrm{Pb}$.
    (a) and (b) compare the NR energy levels with the centroids
    of the corresponding LF energy levels for the same principal
    quantum number $n$.
    (c), (d), and (e) compare the NR energy levels at $n=3$, $n=5$,
    and $n=6$, respectively, with the corresponding LF sublevels, illustrating
    the splitting associated with different longitudinal quantum
    numbers $n_z$.
    Here, $D$ denotes the degeneracy of each level, and
    $C$ represents the cumulative number of states.
    Magic number appears red.}
    \label{fig:NR_LF_comparison}

\end{figure}

For Fig.~\ref{fig:NR_LF_comparison}, $\omega_A$ is calculated using the empirical radius input $R_A=\sqrt{5/3}\,r_{\mathrm{ch,rms}}$ in the Appendix's oscillator-frequency prescription, and $\kappa_A=\sqrt{\Lambda\omega_A}$, as specified in Eq.~\eqref{eq:radius_prescription}. The corresponding parameters are listed in Table~\ref{tab:oscillator-parameters}; the spectrum and state counts follow Eqs.~\eqref{eq:LF_Energy} and \eqref{eq:relativistic_cumulative_transverse_spin_half}.

The light-front energy correction is of order $\omega^2/2\Lambda$ ($\kappa^4/\omega^2$). For general nuclei, $\Lambda$ will be in the GeV range while $\kappa$ will be in the MeV range; thus, the shift in the energy spectrum is not obvious, as we can see in Figs.~\ref{fig:NR_LF_comparison}\subref{fig:Ca48_centroid} and \ref{fig:NR_LF_comparison}\subref{fig:Pb208_centroid}. If we consider the non-relativistic limit, where $\Lambda$ is much larger than $\kappa$, then the $\omega^2/2\Lambda$ term vanishes, and the light-front energy is equivalent to the non-relativistic harmonic oscillator energy.


Another observation in each sub-level cluster we can see in Fig.~\ref{fig:NR_LF_comparison}\subref{fig:Ca48_n4} and Fig.~\ref{fig:NR_LF_comparison}\subref{fig:Pb208_n5} is that the magic numbers $50, 82, 126$ appear in these $n_z=0$ sub-levels' cumulative count of states. In the regular non-relativistic shell model, these numbers only appear after   including the spin-orbit interaction. But they are the lowest levels of each cluster, and it is generally impossible by changing the parameters of the current Hamiltonian to make the spacing between $n_z = 0$ and $n_z =1$ larger than the spacing between $n_z > 0$ states to ensure a complete shell closure description as the regular model does using the spin-orbit interaction.
To obtain a full picture, we would still need to include the spin-orbit interaction. But the special treatment in the longitudinal direction makes the total orbital angular momentum $L$ no longer a good quantum number; therefore, more work should be done to introduce the spin-orbit interaction into our model. 

\subsection{Origin of the Light-Front eigenvalue correction}
\label{sec:origin}

The additional correction obtained in Eq.~\eqref{eq:LF_M^2} and Eq.~\eqref{eq:LF_Energy} breaks the degeneracy of the nonrelativistic harmonic oscillator by introducing an explicit dependence on the longitudinal quantum number \(n_z\). Although this result follows analytically from the light-front eigenvalue equation, its physical origin is not immediately apparent, since both the free-invariant mass term and the longitudinal confinement term can potentially contribute to it. To determine the source of this eigenvalue correction, we first apply the Hellmann–Feynman theorem to separate contributions from the above two terms and then examine how the longitudinal spatial structure and the $P^+$ distribution of the light-front eigenstates influence their confinement expectation values.

\subsubsection{Decomposition of the longitudinal invariant-mass operator}

FLM have established that the light-front harmonic oscillator eigenstate can be separated as in Eq.~\eqref{eq:separation}. Building on that, we can separate the longitudinal eigenvalue problem from the complete two-body Schrödinger equation in Eq.~\eqref{eq:Schrodinger_equation}:

\begin{equation}
\begin{gathered}
M^2_\parallel = Q + U_\parallel,\\
Q = \frac{m_1^2}{x}+\frac{m_2^2}{1-x},
\qquad
U_\parallel =
-\frac{\kappa^4}{\Lambda^2}
\frac{d}{dx}
\left[
x(1-x)\frac{d}{dx}
\right]
\label{eq:longitudinal_M^2}
\end{gathered}
\end{equation}
Define $Q$ as the free longitudinal invariant-mass contribution, while $U_\parallel$ represents longitudinal confinement. In comparison with the free mass contribution in the non-relativistic harmonic oscillator eigenvalue equation, the light-front free mass contribution is non-quadratic in the momentum distribution $x$ and the confinement operator has a specific dependence on the momentum distribution; both can alter the resultant eigenvalue.

From the exact eigenvalue spectrum obtained in Eq.~\eqref{eq:LF_M^2}, we can separate the longitudinal contribution to the eigenvalue:
\begin{equation}
\lambda_{n_z}
=
\Lambda^2
+\kappa^2(2n_z+1)
+\frac{\kappa^4}{\Lambda^2}
n_z(n_z+1).
\label{equ:longitudinal_eigenvalue}
\end{equation}

\subsubsection{Hellmann–Feynman analysis of the eigenvalue contribution}

The eigenvalue receives contributions  from both the free and confinement terms.  We apply the Hellmann–Feynman theorem to identify the eigenvalue contribution from each operator to  locate the source of the additional dependence on $n_z$.
The  Hellmann–Feynman theorem states that  for a normalized eigenstate of a Hermitian Hamiltonian, the derivative of the eigenvalue with respect to a parameter gives the expectation value of the corresponding Hamiltonian derivative:
\begin{equation}
\frac{d\lambda_{n_z}}{da}
=
\left\langle X_{n_z}\left|
\frac{\partial H_\parallel}{\partial a}
\right|X_{n_z}\right\rangle.
\label{eq:Hellmann-Feynman}
\end{equation}
Setting the interaction strength $\kappa^4/\Lambda^2$ to be the parameter $a$, since $Q$ is independent of $a$:
\begin{equation}
\langle U_\parallel\rangle_{\mathrm{LF}}
=
a\frac{d\lambda_{n_z}}{da}
\end{equation}
\begin{equation}
\begin{aligned}
\langle U_\parallel\rangle_{\mathrm{LF}}
&=
\kappa^2\left(n_z+\frac12\right)
+\frac{\kappa^4}{\Lambda^2}n_z(n_z+1).
\label{eq:expect_U}
\end{aligned}
\end{equation}
We have determined the eigenvalue contribution from the longitudinal confinement term. Then, the free longitudinal contribution can be determined by subtracting the confinement expectation from the exact eigenvalue:
\begin{equation}
\begin{aligned}
\langle Q\rangle_{\mathrm{LF}}
&=
\lambda_{n_z}
-\langle U_\parallel\rangle_{\mathrm{LF}}\\
&=
\Lambda^2
+\kappa^2\left(n_z+\frac12\right).
\end{aligned}
\end{equation}
\begin{equation}
\langle Q-\Lambda^2\rangle_{\mathrm{LF}}
=
\kappa^2\left(n_z+\frac12\right)
\label{eq:expect_Q}
\end{equation}

For comparison, the NR harmonic oscillator has equal kinetic and potential expectations. Using the relation $\kappa^2=\Lambda\omega$, we obtain
\begin{equation}
\begin{aligned}
\langle Q-\Lambda^2\rangle_{\mathrm{NR}}
&=\kappa^2\left(n_z+\frac12\right),\\
\langle U_\parallel\rangle_{\mathrm{NR}}
&=\kappa^2\left(n_z+\frac12\right).
\end{aligned}
\end{equation}

Compare with Eqs.~\eqref{eq:expect_Q} and \eqref{eq:expect_U}. The corresponding light-front eigenvalue correction is 
\begin{equation}
\begin{aligned}
\Delta\langle Q\rangle&=0,\\
\Delta\langle U_\parallel\rangle
&=
\frac{\kappa^4}{\Lambda^2}
n_z(n_z+1).
\end{aligned}
\end{equation}

This result establishes that the light-front eigenvalue correction arises entirely from the longitudinal confinement operator, while the free mass longitudinal contribution retains the corresponding NR oscillator expectation value.

\subsubsection{Interpretation of the longitudinal confinement correction}

Knowing that the correction comes from the longitudinal confinement operator, we can investigate its physical origin using the longitudinal structure of the light-front eigenstates. In particular, we examine how the relativistic wave function modifies the expectation value of the longitudinal displacement and how the momentum-dependent coefficient of the confinement operator influences this contribution.

Using the longitudinal displacement operator defined in Eq.~\eqref{eq:longitudinal_M^2}, we can write the expectation value of the confinement operator in integral form
\begin{equation}
\begin{aligned}
\langle U_\parallel\rangle_{\mathrm{LF}}
={}&
\frac{\kappa^4}{\Lambda^2}
\int_0^1\frac{dx}{4\pi}
x(1-x)
\left|\frac{dX(x)}{dx}\right|^2.
\label{U_expect_integral}
\end{aligned}
\end{equation}
The physical meaning of this expression follows from the relationship between the longitudinal momentum fraction and its conjugate displacement operator, the Miller-Brodsky variable $\widetilde Z_\zeta^2$, for which the mean-square holographic longitudinal displacement is 
\begin{equation}
\left\langle
\widetilde Z_\zeta^2
\right\rangle_{\mathrm{LF}}
=
\int_0^1\frac{dx}{4\pi}
\left|
\frac{dX(x)}{dx}
\right|^2.
\end{equation}
Then, we can identify that confinement expectations depend on two related quantities: the longitudinal spatial extent encoded in the derivative of $X(x)$ and the momentum-dependent coefficient $x(1-x)$.
To separate these two effects, we compare the exact light-front expectation with its non-relativistic limit. The free longitudinal invariant-mass contribution is minimized at the equilibrium momentum fraction
\begin{equation}
x_0=\frac{m_1}{\Lambda},
\qquad
x_0(1-x_0)
=
\frac{m_1m_2}{\Lambda^2}.
\end{equation}

In the NR limit, the longitudinal momentum distribution becomes concentrated near $x_0$. Under this approximation, the NR longitudinal confinement expectation becomes
\begin{equation}
\langle U_\parallel\rangle_{\mathrm{NR}}
=
\frac{\kappa^4}{\Lambda^2}
x_0(1-x_0)
\left\langle
\widetilde Z_\zeta^2
\right\rangle_{\mathrm{NR}}.
\label{eq:U_para_NR}
\end{equation}
The exact light-front confinement expectation differs from this NR result for two reasons: the spatial structure of the LF eigenstate differs from its NR counterpart, and the LF confinement operator retains the full dependence on the momentum fraction $x$.

Expressing $x(1-x)$ in terms of the equilibrium $x_0$ and subtracting the NR confinement expectation, we obtain
\begin{equation}
\begin{aligned}
\Delta\langle U_\parallel\rangle
={}&
\frac{\kappa^4}{\Lambda^2}
x_0(1-x_0)
\left[
\left\langle
\widetilde Z_\zeta^2
\right\rangle_{\mathrm{LF}}
-
\left\langle
\widetilde Z_\zeta^2
\right\rangle_{\mathrm{NR}}
\right]\\[5pt]
&+
\frac{\kappa^4}{\Lambda^2}
\int_0^1\frac{dx}{4\pi}
\left[
x(1-x)-x_0(1-x_0)
\right]
\left|
\frac{dX(x)}{dx}
\right|^2.
\end{aligned}
\end{equation}

The first term represents the change in confinement expectation associated with the difference in mean-square holographic longitudinal displacement due to the difference between the light-front and non-relativistic harmonic-oscillator eigenstates. The second term describes the effect of restoring the full momentum-dependent weighting of the LF confinement operator while keeping the exact light-front eigenstate fixed. These are two contributions to the same confinement interaction, and we can express their combined contribution as follows:
\begin{equation}
\Delta\langle U_\parallel\rangle
=
\Delta U_{\mathrm{extent}}
+
\Delta U_{\mathrm{weight}}.
\end{equation}
The exact eigenstates given in Eq.~\eqref{eq:lf_wavefunction} allow the signs of these two contributions to be established analytically.
\begin{equation}
\left\langle
\widetilde Z_\zeta^2
\right\rangle_{\mathrm{LF}}
>
\left\langle
\widetilde Z_\zeta^2
\right\rangle_{\mathrm{NR}},\qquad
\Delta U_{\mathrm{weight}}<0.
\end{equation}
Combining the decomposition above with the Hellmann-Feynman result, we find
\begin{equation}
\Delta U_{\mathrm{extent}}
+
\Delta U_{\mathrm{weight}}
=
\frac{\kappa^4}{\Lambda^2}
n_z(n_z+1).
\end{equation}

Therefore, we can see that the momentum-dependent weighting reduces the eigenvalue compared with its non-relativistic limit, but the change in the structure of the $X(x)$ eigenstate overcomes this reduction and makes the final eigenvalue larger than its corresponding non-relativistic limit. The additional longitudinal correction obtained can therefore be understood as the net confinement cost resulting from the modified spatial structure of the exact light-front eigenstates and the momentum-dependent weighting of the longitudinal interaction.


\subsection{Three properties of the Light-Front energy spectrum}

For realistic nuclei, the light-front correction term is generally smaller than the NR energy separation between adjacent clusters with different n. So states with the same n but different $n_z$ form a sub-level cluster.  After this sub-shell cluster description for the light-front energy spectrum is established, we may focus on the structure inside each sub-shell cluster and the relationship between two sub-shell clusters.

\subsubsection{Sub-shell structure inside One Cluster}

Consider the energy spacing between two adjacent sub-shells in one cluster
with the same $n$ and successive values of $n_z$:
\begin{equation}
\begin{aligned}
&E_{\mathrm{LF}}^{(1)}(n,n_z+1)
&=
\frac{\omega^2}{2\Lambda}(n_z+1).
\end{aligned}
\end{equation}
The energy gap between two sub-shells with the same $n$ depends only on the $n_z$ of the lower sub-shell. This is the manifestation of an increase in the free invariant-mass cost with an increase in longitudinal excitation, as mentioned earlier in Section~\ref{sec:rest}.
Then the difference between successive spacings can be expressed as:
\begin{equation}
\Delta E_1
=
E_{\mathrm{LF}}^{(1)}(n,n_z+1)
-
E_{\mathrm{LF}}^{(1)}(n,n_z)
=
\frac{\omega^2}{\Lambda}(n_z+1),
\end{equation}
\begin{equation}
\Delta E_2
=
E_{\mathrm{LF}}^{(1)}(n,n_z+2)
-
E_{\mathrm{LF}}^{(1)}(n,n_z+1)
=
\frac{\omega^2}{\Lambda}(n_z+2).
\end{equation}
Therefore,
\begin{equation}
\begin{aligned}
\Delta E_2-\Delta E_1
&=
\frac{\omega^2}{\Lambda}
\left[
(n_z+2)-(n_z+1)
\right]
\\
&=
\frac{\omega^2}{\Lambda}
.
\end{aligned}
\end{equation}
The successive spacing difference within the same energy level is a constant value .

\subsubsection{Properties of Clusters}

To ensure that each sub-shell cluster gives the same number of states as the non-relativistic energy level with the same n, sum over the degeneracies of all sub-shell levels that share the same n.
The degeneracy comes from the same value of $2n_\perp+m_\ell$, which is
\begin{equation}
2n_\perp+m_\ell=n-n_z.
\end{equation}

Using Eq.~\eqref{eq:transverse_degeneracy} and then substituting the expression for $n$, we obtain the degeneracy $\tilde g_\mathbf{rel}$
\begin{equation}
\tilde g_\mathbf{rel}
=
N_\perp+1
=
2n_\perp+m_\ell+1
=
n-n_z+1.
\end{equation}
Including the degeneracy due to spin $\frac12$ gives
 \begin{equation}
g_\mathbf{rel}
=
2(n-n_z+1)
\end{equation} so that the total number of states of a cluster is given by 
 $\sum_{n_z=0}^{n}
2(n-n_z+1)
=
(n+1)(n+2)
,$
the same number of states as in the non-relativistic energy level with total quantum number n.

In the nonrelativistic isotropic harmonic oscillator model, all states within a given principal shell are degenerate. In the light-front formulation, a single non-relativistic shell is instead resolved into a cluster of closely spaced light-front sub-shells. The splittings among the individual levels describe the redistribution of states within the cluster, but do not, by themselves, indicate whether the cluster as a whole is shifted relative to the corresponding nonrelativistic shell. To distinguish these effects, we calculate the degeneracy-weighted centroid of each light-front cluster. The difference between this centroid and the energy of the corresponding non-relativistic shell measures the overall relativistic shift, whereas the deviations of the individual levels from the centroid quantify the light-front-induced breaking of the degeneracy.
The centroid is given by
\begin{equation}
\begin{aligned}
\overline{E}_{\mathrm{LF}}^{(1)}(n)
&=
\frac{
\displaystyle
\sum_{n_z=0}^{n}
g_{\mathrm{rel}}(N_\perp,n_z)E_{n,n_z}
}{
D_{n,\mathrm{total}}
}
&=
E_{\mathrm{NR}}(n)
+
\frac{\omega^2}{2\Lambda}
\frac{
\displaystyle
\sum_{n_z=0}^{n}
(n-n_z+1)n_z(n_z+1)
}{
\displaystyle
\sum_{n_z=0}^{n}
(n-n_z+1)
}.
\end{aligned}
\end{equation}
This expression can be simplified to 
\begin{equation}
\begin{aligned}
\overline{E}_{\mathrm{LF}}^{(1)}(n)
&=
E_{\mathrm{NR}}(n)
+
\frac{\omega^2}{12\Lambda}n(n+3)
,
\end{aligned}
\end{equation}
so that 
 $\frac{\omega^2}{12\Lambda}n(n+3)$ is the relativistic correction to the average energy (or energy centroid).

Curiously, the light-front energy correction for each sub-cluster increases quadratically with $n$, which means the energy difference between sub-shell clusters is no longer uniform. In addition, the energy difference between two adjacent clusters will increase as $n$ increases.
Moreover, the correction term is on the scale of $\omega^2/12\Lambda$, which indicates that the correction in the centroid decreases faster than the energy correction for each level as the $\kappa^2/\Lambda^3$ ratio increases.


\subsection{Correction from the Higher-Order $E_{\mathrm{LF}}$ Definition}

Our current definition of rest-frame energy in Eq.~\eqref{eq:energy_first_order} is   obtained by    neglecting terms of higher order in $E_\mathbf{INT}$. This approximation is well justified when the excitation energy is small compared with the total rest-mass scale, but its accuracy decreases for higher excited states or lighter systems. We therefore also retain the higher-order kinematic correction by using the exact relation.
This allows us to distinguish genuine modifications of the light-front spectrum from differences introduced solely by the leading-order conversion between $M^2$ and an energy variable.

Now consider the next-order correction. Set $\Delta M^2=M^2-\Lambda^2$
and $E_{\mathrm{LF}}
=\sqrt{M^2}-\Lambda.$  Then
 \begin{equation}
\begin{aligned}
E_{\mathrm{LF}}
 &=
\frac{\Delta M^2}{2\Lambda}
-
\frac{(\Delta M^2)^2}{8\Lambda^3}
+\cdots .
\end{aligned}
\end{equation}

The first-order term of $E_{\mathrm{LF}}$ is given by Eq.~\eqref{eq:energy_first_order}, so that
\begin{equation}
\frac{(\Delta M^2)^2}{8\Lambda^3}
=
\frac{[E^{(1)}_\mathrm{LF}]^2}{2\Lambda}.
\end{equation}
Thus,
\begin{equation}
E_{\mathrm{LF}}
=
E^{(1)}
-
\frac{[E^{(1)}_\mathrm{LF}]^2}{2\Lambda}.
\end{equation}
The first-order approximation is valid if $\frac{\Delta M^2}{\Lambda^2}\ll1$ which means that $(2n+3)\frac{\omega}{\Lambda}
+
n_z(n_z+1)
\left(
\frac{\omega}{\Lambda}
\right)^2
\ll1.$
Given $\Delta M^2
=
\Lambda\omega(2n+3)
+
\omega^2n_z(n_z+1),$
the second-order term is
 \begin{equation}
\begin{aligned}
\frac{(\Delta M^2)^2}{8\Lambda^3}
&=
\frac{\omega^2(2n+3)^2}{8\Lambda} 
+
\frac{
\omega^3(2n+3)n_z(n_z+1)
}{
4\Lambda^2
}
+
\frac{
\omega^4[n_z(n_z+1)]^2
}{
8\Lambda^3
}.
\end{aligned}
\end{equation}

Consider only the first term:
\begin{equation}
\frac{\omega^2(2n+3)^2}{8\Lambda}
\end{equation}
that  is independent of $n_z$, so it does not change the analysis much. Next we examine the  terms  depending on   $n_z$.

Under this definition of rest-frame bonding energy, the relativistic energy correction gets an extra term:
\begin{equation}
E_{\mathrm{LF}}^{(2)}
-
E_{\mathrm{NR}}
=
\frac{\omega^2}{2\Lambda}
\left[
n_z(n_z+1)
-
\left(
n+\frac{3}{2}
\right)^2
\right].
\end{equation}
Now using $n_z=0$ does not reproduce  the NR energy.

More interestingly, at the largest value of $n_z$, namely $n_z=n$, we obtain
\begin{equation}
\begin{aligned}
n(n+1)
-
\left(
n+\frac{3}{2}
\right)^2
&=
-2n-\frac{9}{4}
,
\end{aligned}
\end{equation}
 indicating that all light-front sub-levels have lower energies than $E_{\mathrm{NR}}(n)$.
Within a cluster, the spacing between sub-levels doesn't change because the new correction is independent of $n_z$.
Then, as expected from the light-front energy correction for the center of mass, we obtain
\begin{equation}
\overline{E}_{\mathrm{LF}}^{(2)}
=
E_{\mathrm{NR}}(n)
+
\frac{\omega^2}{2\Lambda}
\left[
\frac{n(n+3)}{6}
-
\left(
n+\frac{3}{2}
\right)^2
\right].
\end{equation}
There is an extra n-dependent term that has the same scale, $\omega^2/2\Lambda$, as the light-front energy correction, which makes the centroid smaller than the non-relativistic energy level.


Next we examine the higher-order correction that depends on $n_z$.
The individual LF level relative to the NR shell is given by 
\begin{equation}
\begin{aligned}
E_{\mathrm{LF}}^{(3)} - E_{\mathrm{NR}}
&=
\frac{\omega^2}{2\Lambda}
\left[
n_z(n_z+1)
-
\left(n+\frac{3}{2}\right)^2
\right]
 -
\frac{\omega^3(2n+3)}{4\Lambda^2}
\,n_z(n_z+1)
-
\frac{\omega^4}{8\Lambda^3}
\left[n_z(n_z+1)\right]^2 .
\end{aligned}
\end{equation}
The neighboring sub-level spacing is given by 
\begin{equation}
\begin{aligned}
&E_{\mathrm{LF}}(n,n_z+1)^{(3)}
-
E_{\mathrm{LF}}(n,n_z)^{(3)}
 =
\frac{\omega^2}{\Lambda}(n_z+1)
-
\frac{\omega^3(2n+3)}{2\Lambda^2}(n_z+1)
-
\frac{\omega^4}{2\Lambda^3}(n_z+1)^3 .
\end{aligned}
\end{equation}
The weighted cluster centroid, using our previous derivation is given by:
\begin{equation}
\begin{aligned}
\overline{E}_{\mathrm{LF}}^{(3)} - E_{\mathrm{NR}}^{(3)}
&=
\frac{\omega^2}{12\Lambda}
\,n(n+3)
-
\frac{\omega^2(2n+3)^2}{8\Lambda}
-
\frac{\omega^3(2n+3)n(n+3)}{24\Lambda^2}
-
\frac{\omega^4 n(n+3)(n^2+3n+1)}
{120\Lambda^3}.
\end{aligned}
\end{equation}
This completes the discussion of the relativistic corrections to the energy levels.

\section{Form factor}
\label{sec:ff}

In previous sections, we have elaborated on the features of the $M^2$ eigenvalues and their applications to the nuclear shell model (or independent-particle model). We now focus on the electromagnetic  form factor. This  has received the most attention because it is used to determine the charge density of the nucleus.  See {\it e.g.}~\cite{Frosch:1968zz}.

\subsection{Form factor in the light-front formalism}

For a spin-zero system, Lorentz covariance and current conservation allow the electromagnetic current matrix element to be written in terms of a single scalar form factor.
\begin{equation}
\left\langle P' \middle| J^\mu(0) \middle| P \right\rangle
=
(P'+P)^\mu F(Q^2),
\qquad
Q^2=-q^2,
\qquad
q=P'-P .
\end{equation}
The function $F(Q^2)$ is   relativistically invariant. To evaluate it in light-front dynamics, it is convenient to choose the Drell--Yan frame~\cite{Miller_2009} with 
$q^+=0,
\,
Q^2=\mathbf q_\perp^2 .$
Since the value of $P^+$ is not  changed by the scattering process,   the form factor is 
computed  using  
\begin{equation}
F(Q^2)
=
\frac{1}{2P^+}
\left\langle P' \middle|J^+(0)\middle|P\right\rangle .
\end{equation}

For a two-body light-front state, let \(x\) denote the fraction of total plus momentum carried by the struck constituent. Since $q^+=0$, $x$ is unchanged by the interaction, while the struck constituent's relative transverse momentum changes according to
\begin{equation}
\mathbf k'_\perp
=
\mathbf k_\perp
+
(1-x)\mathbf q_\perp .
\end{equation}
Then the relativistic form factor is given by
\begin{equation}
F(Q^2)
=
\int_0^1
\frac{dx}{4\pi x(1-x)}
\int
\frac{d^2\mathbf k_\perp}{(2\pi)^2}
\,
\psi^*
\!\left(
x,\mathbf k_\perp+(1-x)\mathbf q_\perp
\right)
\psi(x,\mathbf k_\perp).
\end{equation}

\subsection{LF and NR form factor comparison}

The non-relativistic form factor is calculated using  a Fourier transform of the charge distribution of the nucleus.
\begin{equation}
F_{\mathrm{NR}}(Q^2)
=
\int d^3\mathbf r\,
|\psi_{\mathrm{NR}}(\mathbf r)|^2
\exp\!\left[
i\,\frac{m_2}{m_1+m_2}\,
\mathbf q\cdot\mathbf r
\right],
\qquad
Q^2=\mathbf q^2 .
\end{equation}

For the numerical calculations, we retain the oscillator-frequency prescription in the Appendix and use two radius inputs. In the empirical prescription, the evaluated rms charge radii $r_{\mathrm{ch,rms}}=3.4771(20)\,\mathrm{fm}$ for $^{48}\mathrm{Ca}$ and $5.5012(13)\,\mathrm{fm}$ for $^{208}\mathrm{Pb}$~\cite{AngeliMarinova2013} are converted to equivalent uniform-sphere radii,
\begin{equation}
R_A=\sqrt{\frac{5}{3}}\,r_{\mathrm{ch,rms}},
\qquad
\kappa_A=\sqrt{\Lambda\omega_A}.
\label{eq:radius_prescription}
\end{equation}
The second prescription uses $R_A=1.2A^{1/3}\,\mathrm{fm}$ in the same oscillator-frequency expression. These are radius-input prescriptions for the oscillator parameters; the empirical prescription is not a direct fit to the LF form-factor slope. Table~\ref{tab:oscillator-parameters} lists the radius inputs and the resulting parameters for both methods.

\begin{figure*}[t]
    \centering

    \begin{subfigure}{0.48\textwidth}
        \centering
        \includegraphics[width=\linewidth]{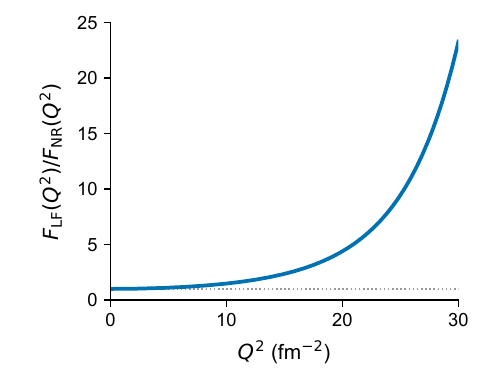}
        \caption{}
        \label{fig:Ca48_method2}
    \end{subfigure}
    \hfill
    \begin{subfigure}{0.48\textwidth}
        \centering
        \includegraphics[width=\linewidth]{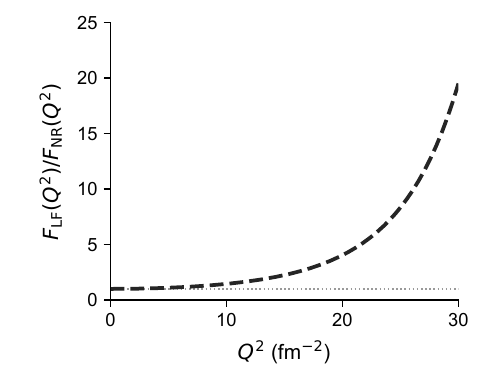}
        \caption{}
        \label{fig:Ca48_method3}
    \end{subfigure}

    \vspace{0.3cm}

    \begin{subfigure}{0.48\textwidth}
        \centering
        \includegraphics[width=\linewidth]{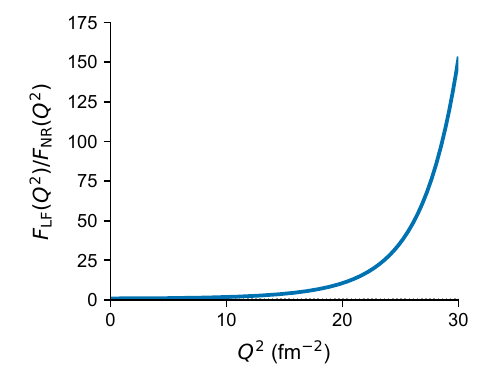}
        \caption{}
        \label{fig:Pb208_method2}
    \end{subfigure}
    \hfill
    \begin{subfigure}{0.48\textwidth}
        \centering
        \includegraphics[width=\linewidth]{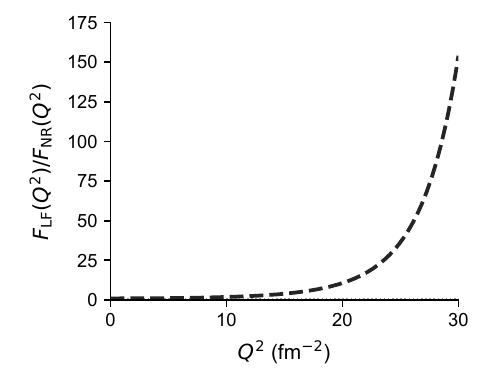}
        \caption{}
        \label{fig:Pb208_method3}
    \end{subfigure}

    \caption{
    Comparison of form factor ratios for
    $^{48}\mathrm{Ca}$ and $^{208}\mathrm{Pb}$
    using two different methods for calculating $\kappa$.
    (a) and (b) show the results for $^{48}\mathrm{Ca}$
    using the empirical charge-radius-based method
    and the nuclear-radius approximation, respectively.
    (c) and (d) show the corresponding results
    for $^{208}\mathrm{Pb}$.
    }
    \label{fig:form_factor_comparison}

\end{figure*}

The frequency calculation uses $m_1=0.93827\,\mathrm{GeV}$ and $m_2=(A-1)u$, with $u=0.93149\,\mathrm{GeV}$. To reproduce the original numerical form-factor prescription, Fig.~\ref{fig:form_factor_comparison} uses the approximate parameters $\alpha_A=2Am_1^2/\kappa_A^2$, $\beta_A=(A-1)\alpha_A$, and the NR factor $\exp[-(A-1)Q^2/(4\kappa_A^2)]$ in natural units. The spectrum in Fig.~\ref{fig:NR_LF_comparison} retains the general mass definitions in Eq.~\eqref{eq:alpha_beta}. The form-factor ratios are displayed on linear vertical axes.

We can directly observe in Figs.~\ref{fig:form_factor_comparison}\subref{fig:Ca48_method2}, \ref{fig:form_factor_comparison}\subref{fig:Ca48_method3}, \ref{fig:form_factor_comparison}\subref{fig:Pb208_method2}, and \ref{fig:form_factor_comparison}\subref{fig:Pb208_method3} that, in either method of harmonic-potential modeling, the non-relativistic form factor is a few orders of magnitude smaller than the light-front form factor; the dispersion between the two treatments increases with $q^2$ and becomes considerable at large $q^2$. In the current region of interest, the $^{48}\mathrm{Ca}$ form factor ratio can extend to 20. The difference for $^{208}\mathrm{Pb}$ is much more significant as it can extend to 140. ${}^{208}\mathrm{Pb}$ has a larger form factor ratio than ${}^{48}\mathrm{Ca}$ across the two treatments;
it is generally true that in the current model, isotopes of larger mass have a larger dispersion in non-relativistic and light-front form factors than do isotopes of smaller mass.

The substantial difference between the form factors of the two isotopes indicates that the relativistic correction is important in describing  nuclear structure,  even though corrections to the energy are small.    

\section{Conclusion}
\label{sec:end}

We investigated the applications of the three-dimensional, boost-invariant light-front harmonic oscillator to nuclear structure and established how it modifies the non-relativistic description. Based on a two-body description of a nucleon and the remaining spectator nucleus, we suggest a relativistic shell model with a light-front invariant-mass-squared spectrum and analyze the relativistic correction to the form factor for specific isotopes.

The study of the light-front invariant-mass-squared spectrum suggests that the relativistic correction depends on longitudinal excitation, which breaks the degeneracy of the regular non-relativistic three-dimensional harmonic-oscillator shells. We analyzed characteristic features resulting from this broken degeneracy in the light-front shell model. Furthermore, using the Hellmann–Feynman theorem, we indicate that the eigenvalue (energy or invariant mass) correction term originates from the longitudinal confinement operator rather than the free invariant-mass term. This correction can be understood through the balance of effects from the modified longitudinal spatial structure of the light-front eigenstates and the momentum-fraction dependence of the confining interaction.

We calculated the electromagnetic form factors for $^{48}\mathrm{Ca}$ and $^{208}\mathrm{Pb}$ using both the non-relativistic and light-front harmonic-oscillator models and compared their results. We found that the non-relativistic treatment underestimates the form factors relative to the light-front calculation, with the discrepancy increasing with momentum transfer and becoming more pronounced for isotopes with larger nucleon numbers. These results indicate that, although the relativistic corrections to the energy spectrum are relatively small, modifications of the bound-state wave functions can lead to substantial differences in the predicted electromagnetic structure of nuclei.

Overall, our work demonstrates an early  application of light-front dynamics to nuclear structure and illustrates how relativistic effects modify the conventional non-relativistic description by examining both the eigenvalues and eigenstates of the harmonic oscillator. An important direction for future work is to incorporate spin–orbit interactions into the light-front harmonic-oscillator framework, extending the present model toward a complete relativistic nuclear shell model. Such a model could be the basis for more detailed calculations including correlations.

\begin{acknowledgments}
  AF was support by
  the Center for Nuclear Femtography,
  operated by the Southeastern Universities Research Association
  in Washington, D.C.\ under an appropriation from the Commonwealth of Virginia;
  and by the U.S.\ Department of Energy,
  Office of Science, Office of Nuclear Physics under Contract No.\ 89243126CSC000213.
  GAM is partially supported by the U.S.\ Department of Energy,
  Office of Science, Office of Nuclear Physics,
  under contract number DE-SC0026252.
\end{acknowledgments}


\appendix

\section{Correction to the shell-model calculation of interaction strength}

The general shell-model treatment of a nucleus calculates the harmonic-oscillation frequency (and then the interaction strength) using
\begin{align}
\hbar \omega_0
&=
\frac{1}{A}
\frac{\hbar^2}{\mu \langle r^2 \rangle}
\frac{1}{2}
\left(\frac{3}{2}A\right)^{4/3}
=
\frac{5}{4}
\left(\frac{3}{2}\right)^{1/3}
\frac{\hbar^2}{\mu r_0^2}
A^{-1/3}
\approx
41 A^{-1/3}\,\mathrm{MeV}.
\end{align}
where $\mu$ is the reduced mass,   defined as $\mu = (A-1)m_p/A$.
There are three different approximations in this   harmonic-oscillator
derivation:
\begin{enumerate}
    \item The proton and neutron numbers are assumed to be equal,
    \begin{equation}
    Z=N_n=\frac{A}{2},
    \end{equation}
    where $A=Z+N_n$.

    \item The proton and neutron shell occupations are assumed to be large,
    so the large-$N_{\max}$ approximation can be used.
    \item The proton and neutron numbers always correspond to completely filled shells.
\end{enumerate}
Here, we to analyze the corrections to the first two assumptions.

\subsection{Correction from unequal proton and neutron numbers}

In this section, the large-number approximation is still kept. The only
assumption that is removed is that $Z=N_n.$
 For protons, the number of particles in completely filled harmonic-oscillator
shells is
\begin{equation}
Z
=
\frac{1}{3}
\left(N_{\max}^{(p)}+1\right)
\left(N_{\max}^{(p)}+2\right)
\left(N_{\max}^{(p)}+3\right).
\end{equation}
This can be rewritten as
\begin{equation}
3Z
=
\left(N_{\max}^{(p)}+2\right)^3
-
\left(N_{\max}^{(p)}+2\right).
\end{equation}

Under the large-number approximation,
\begin{equation}
N_{\max}^{(p)}+2
\approx
(3Z)^{1/3}.
\end{equation}

Similarly, for neutrons,
\begin{equation}
N_{\max}^{(n)}+2
\approx
(3N_n)^{1/3}.
\end{equation}
In the textbook~\cite{Wong1998}, the expectation value of the squared radius for a given number of nucleons based on the average squared radius is:
\begin{equation}
\left\langle R^2 \right\rangle
=
\frac{1}{A}
\frac{\hbar}{\mu\omega_0}
\frac{1}{4}
\left(N_{\max}+1\right)
\left(N_{\max}+2\right)^2
\left(N_{\max}+3\right)
\end{equation}

For separate proton and neutron numbers, the oscillator frequency is
\begin{equation}
\hbar\omega_0
=
\frac{3\hbar^2}
{4A\mu\langle R^2\rangle}
\left[
Z\left(N_{\max}^{(p)}+2\right)
+
N_n\left(N_{\max}^{(n)}+2\right)
\right].
\end{equation}

Using the large-number approximation gives
\begin{equation}
\hbar\omega_0(Z,N_n)
\approx
\frac{3^{4/3}\hbar^2}
{4A\mu\langle R^2\rangle}
\left(
Z^{4/3}+N_n^{4/3}
\right).
\label{eq:omega_unequal_large}
\end{equation}

The equal-proton-neutron approximation corresponds to
\begin{equation}
Z=N_n=\frac{A}{2}.
\end{equation}

Therefore,
\begin{equation}
\hbar\omega_0(Z=N_n)
=
\frac{3^{4/3}\hbar^2}
{4A\mu\langle R^2\rangle}
\,2
\left(\frac{A}{2}\right)^{4/3}.
\label{eq:omega_equal_large}
\end{equation}

To determine the correction caused by $Z\neq N_n$, we characterize the deviation as the ratio between the proton-neutron number difference and the total nucleon number $(N_n-Z)/A$.
Write $Z$ and $N_n$ in terms of this number-difference ratio.
\begin{equation}
\begin{aligned}
Z^{4/3}+N_n^{4/3}
=
\left(\frac{A}{2}\right)^{4/3}
\Bigg[
&
\left(
1-\frac{N_n-Z}{A}
\right)^{4/3}
 +
\left(
1+\frac{N_n-Z}{A}
\right)^{4/3}
\Bigg].
\end{aligned}
\end{equation}

Using the binomial expansions
\begin{equation}
Z^{4/3}+N_n^{4/3}
=
2
\left(\frac{A}{2}\right)^{4/3}
\left[
1+
\frac{2}{9}
\left(
\frac{N_n-Z}{A}
\right)^2
+\cdots
\right].
\end{equation}

Substituting this result into Eq.~\eqref{eq:omega_unequal_large} gives
\begin{equation}
\hbar\omega_0(Z,N_n)
\approx
\hbar\omega_0(Z=N_n)
\left[
1+
\frac{2}{9}
\left(
\frac{N_n-Z}{A}
\right)^2
\right].
\end{equation}

Hence, the leading correction caused by treating the proton and neutron
numbers separately is
\begin{equation}
\Delta(\hbar\omega_0)_{Z\neq N_n}
=
\hbar\omega_0(Z=N_n)
=\frac{2}{9}
\left(
\frac{N_n-Z}{A}
\right)^2.
\end{equation}
Equivalently, the fractional correction is
\begin{equation}
\frac{\Delta\omega_0}{\omega_0}
=
\frac{2}{9}
\left(
\frac{N_n-Z}{A}
\right)^2.
\label{eq:unequal_fractional_correction}
\end{equation}
Therefore, there is no correction linear in $N_n-Z$. The leading-order correction
from proton-neutron asymmetry is quadratic in $(N_n-Z)/A$.

\subsection{Correction from removing the large-number approximation}

We now keep the proton and neutron numbers separate from the beginning.
The comparison is therefore between the result with separate $Z$ and $N_n$
under the large-number approximation and the result without the
large-number approximation.

For protons, the exact relation is
\begin{equation}
3Z
=
\left(N_{\max}^{(p)}+2\right)^3
-
\left(N_{\max}^{(p)}+2\right).
\label{eq:proton_exact_count}
\end{equation}
Similarly, for neutrons,
\begin{equation}
3N_n
=
\left(N_{\max}^{(n)}+2\right)^3
-
\left(N_{\max}^{(n)}+2\right).
\label{eq:neutron_exact_count}
\end{equation}

The large-number approximation neglects the second term in each equation,
giving
\begin{equation}
N_{\max}^{(p)}+2
\approx
(3Z)^{1/3},\,N_{\max}^{(n)}+2
\approx
(3N_n)^{1/3}.
\end{equation}
 The corresponding large-number oscillator frequency is
\begin{equation}
\hbar\omega_0^{\mathrm{large}}
=
\frac{3\hbar^2}
{4A\mu\langle R^2\rangle}
\left[
Z(3Z)^{1/3}
+
N_n(3N_n)^{1/3}
\right].
\label{eq:omega_large_separate}
\end{equation}
To find the leading correction, define
\begin{equation}
N_{\max}^{(p)}+2
=
(3Z)^{1/3}
+
\epsilon_p.
\end{equation}
where $\epsilon_p$ is used to determine the correction term.
Substituting this into Eq.~\eqref{eq:proton_exact_count} and keeping only first-order terms in $\epsilon_p$ gives
\begin{equation}
\epsilon_p
\approx
\frac{1}
{3(3Z)^{1/3}},
\end{equation}
and hence
\begin{equation}
N_{\max}^{(p)}+2
\approx
(3Z)^{1/3}
+
\frac{1}
{3(3Z)^{1/3}}.
\label{eq:proton_finite_correction}
\end{equation}
The same calculation for neutrons gives
\begin{equation}
N_{\max}^{(n)}+2
\approx
(3N_n)^{1/3}
+
\frac{1}
{3(3N_n)^{1/3}}.
\label{eq:neutron_finite_correction}
\end{equation}
The exact expression for the oscillator frequency is then
\begin{equation}
\hbar\omega_0
=
\frac{3\hbar^2}
{4A\mu\langle R^2\rangle}
\left[
Z\left(N_{\max}^{(p)}+2\right)
+
N_n\left(N_{\max}^{(n)}+2\right)
\right].
\end{equation}
Substituting Eqs.~\eqref{eq:proton_finite_correction} and
\eqref{eq:neutron_finite_correction} gives
\begin{equation}
\begin{aligned}
\hbar\omega_0
\approx{}&
\frac{3\hbar^2}
{4A\mu\langle R^2\rangle}
\left[
Z(3Z)^{1/3}
+
N_n(3N_n)^{1/3}
\right]
+
\frac{3\hbar^2}
{4A\mu\langle R^2\rangle}
\left[
\frac{Z}{3(3Z)^{1/3}}
+
\frac{N_n}{3(3N_n)^{1/3}}
\right].
\end{aligned}
\end{equation}
The first line is precisely the separate-proton-neutron large-number result
in Eq.~\eqref{eq:omega_large_separate}. Therefore, the leading correction
caused by removing the large-number approximation is
\begin{equation}
\Delta(\hbar\omega_0)_{\mathrm{finite}}
=
\frac{3\hbar^2}
{4A\mu\langle R^2\rangle}
\left[
\frac{Z}{3(3Z)^{1/3}}
+
\frac{N_n}{3(3N_n)^{1/3}}
\right].
\end{equation}

This can be simplified to
\begin{equation}
\Delta(\hbar\omega_0)_{\mathrm{finite}}
=
\frac{\hbar^2}
{4A\mu\langle R^2\rangle\,3^{1/3}}
\left(
Z^{2/3}+N_n^{2/3}
\right).
\label{eq:finite_number_correction}
\end{equation}

Dividing by the large-number result gives the fractional correction
\begin{equation}
\frac{
\Delta\omega_0{}_{\mathrm{finite}}
}{
\omega_0^{\mathrm{large}}
}
=
\frac{
Z^{2/3}+N_n^{2/3}
}{
3^{5/3}
\left(
Z^{4/3}+N_n^{4/3}
\right)
}.
\label{eq:finite_fractional_correction}
\end{equation}

\subsection{Summary}

For $^{48}\mathrm{Ca}$, the correction due to proton--neutron asymmetry is approximately $0.62\%$, while the correction from removing the large-occupation-number approximation is approximately $1.91\%$. For $^{208}\mathrm{Pb}$, the corresponding corrections are approximately $0.99\%$ and $0.71\%$, respectively. Therefore, both corrections are at the percent level for the nuclei considered in this work and should not be neglected in the numerical calculation.

For the numerical prescriptions, $\omega_A$ and $\kappa_A$ are obtained using the expression below with either the empirical input $R_A=\sqrt{5/3}\,r_{\mathrm{ch,rms}}$ or the nuclear-radius approximation $R_A=1.2A^{1/3}\,\mathrm{fm}$, as specified in Eq.~\eqref{eq:radius_prescription}. The numerical expression is retained in its original form.

\[
\omega_A
=
\frac{3(\hbar c)^2 
[ZC(Z)+NC(N)]
}
{4A\mu R_A^2}
\]

\[
C(q)
=
\left(
\frac{3q}{2}
+
\sqrt{\frac{9q^2}{4}-\frac{1}{27}}
\right)^{1/3}
+
\left(
\frac{3q}{2}
-
\sqrt{\frac{9q^2}{4}-\frac{1}{27}}
\right)^{1/3}
\]

\begin{table}[tb]
  \centering
  \caption{Calculated oscillator parameters for $^{48}\mathrm{Ca}$ and
  $^{208}\mathrm{Pb}$.}
  \label{tab:oscillator-parameters}
  {\small\setlength{\tabcolsep}{4pt}
  \begin{tabular}{lccccccc}
    \toprule
    & & \multicolumn{3}{c}{Empirical radius conversion}
      & \multicolumn{3}{c}{Nuclear-radius approximation} \\
    \cmidrule(lr){3-5}\cmidrule(lr){6-8}
    Nucleus & $r_{\mathrm{ch,rms}}$ (fm)
      & $R_A$ (fm) & $\omega_A$ (MeV) & $\kappa_A$ (GeV)
      & $R_A$ (fm) & $\omega_A$ (MeV) & $\kappa_A$ (GeV) \\
    \midrule
    $^{48}\mathrm{Ca}$  & 3.4771(20) & 4.4889(26) & 6.7302(77) & 0.54860(32) & 4.36109 & 7.13057 & 0.564683 \\
    $^{208}\mathrm{Pb}$ & 5.5012(13) & 7.1020(17) & 4.2780(20) & 0.91043(22) & 7.10999 & 4.26841 & 0.909413 \\
    \bottomrule
  \end{tabular}}
  \par\smallskip
  {\footnotesize Experimental rms charge radii are from Ref.~\cite{AngeliMarinova2013}.
  The empirical radius input is $R_A=\sqrt{5/3}\,r_{\mathrm{ch,rms}}$; the second input is $R_A=1.2A^{1/3}\,\mathrm{fm}$.
  Parentheses give uncertainties in the last quoted digits. The uncertainties in
  the empirical $R_A$, $\omega_A$, and $\kappa_A$ are propagated from the measured radius uncertainties
  only; no uncertainty was specified for the nuclear-radius approximation.}
\end{table}

\bibliography{reference}

\begin{thebibliography}{28}%
\makeatletter
\providecommand \@ifxundefined [1]{%
 \@ifx{#1\undefined}
}%
\providecommand \@ifnum [1]{%
 \ifnum #1\expandafter \@firstoftwo
 \else \expandafter \@secondoftwo
 \fi
}%
\providecommand \@ifx [1]{%
 \ifx #1\expandafter \@firstoftwo
 \else \expandafter \@secondoftwo
 \fi
}%
\providecommand \natexlab [1]{#1}%
\providecommand \enquote  [1]{``#1''}%
\providecommand \bibnamefont  [1]{#1}%
\providecommand \bibfnamefont [1]{#1}%
\providecommand \citenamefont [1]{#1}%
\providecommand \href@noop [0]{\@secondoftwo}%
\providecommand \href [0]{\begingroup \@sanitize@url \@href}%
\providecommand \@href[1]{\@@startlink{#1}\@@href}%
\providecommand \@@href[1]{\endgroup#1\@@endlink}%
\providecommand \@sanitize@url [0]{\catcode `\\12\catcode `\$12\catcode `\&12\catcode `\#12\catcode `\^12\catcode `\_12\catcode `\%12\relax}%
\providecommand \@@startlink[1]{}%
\providecommand \@@endlink[0]{}%
\providecommand \url  [0]{\begingroup\@sanitize@url \@url }%
\providecommand \@url [1]{\endgroup\@href {#1}{\urlprefix }}%
\providecommand \urlprefix  [0]{URL }%
\providecommand \Eprint [0]{\href }%
\providecommand \doibase [0]{https://doi.org/}%
\providecommand \selectlanguage [0]{\@gobble}%
\providecommand \bibinfo  [0]{\@secondoftwo}%
\providecommand \bibfield  [0]{\@secondoftwo}%
\providecommand \translation [1]{[#1]}%
\providecommand \BibitemOpen [0]{}%
\providecommand \bibitemStop [0]{}%
\providecommand \bibitemNoStop [0]{.\EOS\space}%
\providecommand \EOS [0]{\spacefactor3000\relax}%
\providecommand \BibitemShut  [1]{\csname bibitem#1\endcsname}%
\let\auto@bib@innerbib\@empty
\bibitem [{\citenamefont {Dirac}(1949)}]{Dirac:1949cp}%
  \BibitemOpen
  \bibfield  {author} {\bibinfo {author} {\bibfnamefont {P.~A.~M.}\ \bibnamefont {Dirac}},\ }\bibfield  {title} {\bibinfo {title} {{Forms of Relativistic Dynamics}},\ }\href {https://doi.org/10.1103/RevModPhys.21.392} {\bibfield  {journal} {\bibinfo  {journal} {Rev. Mod. Phys.}\ }\textbf {\bibinfo {volume} {21}},\ \bibinfo {pages} {392} (\bibinfo {year} {1949})}\BibitemShut {NoStop}%
\bibitem [{\citenamefont {Newton}\ and\ \citenamefont {Wigner}(1949)}]{Newton:1949cq}%
  \BibitemOpen
  \bibfield  {author} {\bibinfo {author} {\bibfnamefont {T.~D.}\ \bibnamefont {Newton}}\ and\ \bibinfo {author} {\bibfnamefont {E.~P.}\ \bibnamefont {Wigner}},\ }\bibfield  {title} {\bibinfo {title} {{Localized States for Elementary Systems}},\ }\href {https://doi.org/10.1103/RevModPhys.21.400} {\bibfield  {journal} {\bibinfo  {journal} {Rev. Mod. Phys.}\ }\textbf {\bibinfo {volume} {21}},\ \bibinfo {pages} {400} (\bibinfo {year} {1949})}\BibitemShut {NoStop}%
\bibitem [{\citenamefont {Miller}(2025)}]{miller2025impossibility}%
  \BibitemOpen
  \bibfield  {author} {\bibinfo {author} {\bibfnamefont {G.~A.}\ \bibnamefont {Miller}},\ }\bibfield  {title} {\bibinfo {title} {{Impossibility of obtaining time-independent, three-dimensional, spherically symmetric densities of confined systems of relativistically moving constituents}},\ }\href {https://doi.org/10.1103/jkn7-4fzj} {\bibfield  {journal} {\bibinfo  {journal} {Phys. Rev. C}\ }\textbf {\bibinfo {volume} {112}},\ \bibinfo {pages} {045204} (\bibinfo {year} {2025})},\ \Eprint {https://arxiv.org/abs/2507.14388} {arXiv:2507.14388 [hep-ph]} \BibitemShut {NoStop}%
\bibitem [{\citenamefont {Feynman}\ \emph {et~al.}(1971)\citenamefont {Feynman}, \citenamefont {Kislinger},\ and\ \citenamefont {Ravndal}}]{Feynman:1971wr}%
  \BibitemOpen
  \bibfield  {author} {\bibinfo {author} {\bibfnamefont {R.~P.}\ \bibnamefont {Feynman}}, \bibinfo {author} {\bibfnamefont {M.}~\bibnamefont {Kislinger}},\ and\ \bibinfo {author} {\bibfnamefont {F.}~\bibnamefont {Ravndal}},\ }\bibfield  {title} {\bibinfo {title} {{Current matrix elements from a relativistic quark model}},\ }\href {https://doi.org/10.1103/PhysRevD.3.2706} {\bibfield  {journal} {\bibinfo  {journal} {Phys. Rev. D}\ }\textbf {\bibinfo {volume} {3}},\ \bibinfo {pages} {2706} (\bibinfo {year} {1971})}\BibitemShut {NoStop}%
\bibitem [{\citenamefont {Leutwyler}\ and\ \citenamefont {Stern}(1978)}]{Leutwyler:1977vy}%
  \BibitemOpen
  \bibfield  {author} {\bibinfo {author} {\bibfnamefont {H.}~\bibnamefont {Leutwyler}}\ and\ \bibinfo {author} {\bibfnamefont {J.}~\bibnamefont {Stern}},\ }\bibfield  {title} {\bibinfo {title} {{Relativistic Dynamics on a Null Plane}},\ }\href {https://doi.org/10.1016/0003-4916(78)90082-9} {\bibfield  {journal} {\bibinfo  {journal} {Annals Phys.}\ }\textbf {\bibinfo {volume} {112}},\ \bibinfo {pages} {94} (\bibinfo {year} {1978})}\BibitemShut {NoStop}%
\bibitem [{\citenamefont {Kogut}\ and\ \citenamefont {Soper}(1970)}]{Kogut:1969xa}%
  \BibitemOpen
  \bibfield  {author} {\bibinfo {author} {\bibfnamefont {J.~B.}\ \bibnamefont {Kogut}}\ and\ \bibinfo {author} {\bibfnamefont {D.~E.}\ \bibnamefont {Soper}},\ }\bibfield  {title} {\bibinfo {title} {{Quantum Electrodynamics in the Infinite Momentum Frame}},\ }\href {https://doi.org/10.1103/PhysRevD.1.2901} {\bibfield  {journal} {\bibinfo  {journal} {Phys. Rev. D}\ }\textbf {\bibinfo {volume} {1}},\ \bibinfo {pages} {2901} (\bibinfo {year} {1970})}\BibitemShut {NoStop}%
\bibitem [{\citenamefont {Kogut}\ and\ \citenamefont {Susskind}(1973)}]{Kogut:1972di}%
  \BibitemOpen
  \bibfield  {author} {\bibinfo {author} {\bibfnamefont {J.~B.}\ \bibnamefont {Kogut}}\ and\ \bibinfo {author} {\bibfnamefont {L.}~\bibnamefont {Susskind}},\ }\bibfield  {title} {\bibinfo {title} {{The Parton picture of elementary particles}},\ }\href {https://doi.org/10.1016/0370-1573(73)90009-4} {\bibfield  {journal} {\bibinfo  {journal} {Phys. Rept.}\ }\textbf {\bibinfo {volume} {8}},\ \bibinfo {pages} {75} (\bibinfo {year} {1973})}\BibitemShut {NoStop}%
\bibitem [{\citenamefont {Brodsky}\ \emph {et~al.}(1998)\citenamefont {Brodsky}, \citenamefont {Pauli},\ and\ \citenamefont {Pinsky}}]{Brodsky:1997de}%
  \BibitemOpen
  \bibfield  {author} {\bibinfo {author} {\bibfnamefont {S.~J.}\ \bibnamefont {Brodsky}}, \bibinfo {author} {\bibfnamefont {H.-C.}\ \bibnamefont {Pauli}},\ and\ \bibinfo {author} {\bibfnamefont {S.~S.}\ \bibnamefont {Pinsky}},\ }\bibfield  {title} {\bibinfo {title} {{Quantum chromodynamics and other field theories on the light cone}},\ }\href {https://doi.org/10.1016/S0370-1573(97)00089-6} {\bibfield  {journal} {\bibinfo  {journal} {Phys. Rept.}\ }\textbf {\bibinfo {volume} {301}},\ \bibinfo {pages} {299} (\bibinfo {year} {1998})},\ \Eprint {https://arxiv.org/abs/hep-ph/9705477} {arXiv:hep-ph/9705477} \BibitemShut {NoStop}%
\bibitem [{\citenamefont {Miller}(2000)}]{Miller:2000kv}%
  \BibitemOpen
  \bibfield  {author} {\bibinfo {author} {\bibfnamefont {G.~A.}\ \bibnamefont {Miller}},\ }\bibfield  {title} {\bibinfo {title} {{Light front quantization: A Technique for relativistic and realistic nuclear physics}},\ }\href {https://doi.org/10.1016/S0146-6410(00)00103-4} {\bibfield  {journal} {\bibinfo  {journal} {Prog. Part. Nucl. Phys.}\ }\textbf {\bibinfo {volume} {45}},\ \bibinfo {pages} {83} (\bibinfo {year} {2000})},\ \Eprint {https://arxiv.org/abs/nucl-th/0002059} {arXiv:nucl-th/0002059} \BibitemShut {NoStop}%
\bibitem [{\citenamefont {Bakker}\ \emph {et~al.}(1979)\citenamefont {Bakker}, \citenamefont {Kondratyuk},\ and\ \citenamefont {Terentev}}]{Bakker:1979eg}%
  \BibitemOpen
  \bibfield  {author} {\bibinfo {author} {\bibfnamefont {B.~L.~G.}\ \bibnamefont {Bakker}}, \bibinfo {author} {\bibfnamefont {L.~A.}\ \bibnamefont {Kondratyuk}},\ and\ \bibinfo {author} {\bibfnamefont {M.~V.}\ \bibnamefont {Terentev}},\ }\bibfield  {title} {\bibinfo {title} {{On The Formulation of Two-Body and Three-body Relativistic Equations Employing Light Front Dynamics}},\ }\href {https://doi.org/10.1016/0550-3213(79)90179-2} {\bibfield  {journal} {\bibinfo  {journal} {Nucl. Phys. B}\ }\textbf {\bibinfo {volume} {158}},\ \bibinfo {pages} {497} (\bibinfo {year} {1979})}\BibitemShut {NoStop}%
\bibitem [{\citenamefont {Carbonell}\ \emph {et~al.}(1998)\citenamefont {Carbonell}, \citenamefont {Desplanques}, \citenamefont {Karmanov},\ and\ \citenamefont {Mathiot}}]{Carbonell:1998rj}%
  \BibitemOpen
  \bibfield  {author} {\bibinfo {author} {\bibfnamefont {J.}~\bibnamefont {Carbonell}}, \bibinfo {author} {\bibfnamefont {B.}~\bibnamefont {Desplanques}}, \bibinfo {author} {\bibfnamefont {V.~A.}\ \bibnamefont {Karmanov}},\ and\ \bibinfo {author} {\bibfnamefont {J.~F.}\ \bibnamefont {Mathiot}},\ }\bibfield  {title} {\bibinfo {title} {{Explicitly covariant light front dynamics and relativistic few body systems}},\ }\href {https://doi.org/10.1016/S0370-1573(97)00090-2} {\bibfield  {journal} {\bibinfo  {journal} {Phys. Rept.}\ }\textbf {\bibinfo {volume} {300}},\ \bibinfo {pages} {215} (\bibinfo {year} {1998})},\ \Eprint {https://arxiv.org/abs/nucl-th/9804029} {arXiv:nucl-th/9804029} \BibitemShut {NoStop}%
\bibitem [{\citenamefont {Blunden}\ \emph {et~al.}(1999{\natexlab{a}})\citenamefont {Blunden}, \citenamefont {Burkardt},\ and\ \citenamefont {Miller}}]{Blunden:1999gq}%
  \BibitemOpen
  \bibfield  {author} {\bibinfo {author} {\bibfnamefont {P.~G.}\ \bibnamefont {Blunden}}, \bibinfo {author} {\bibfnamefont {M.}~\bibnamefont {Burkardt}},\ and\ \bibinfo {author} {\bibfnamefont {G.~A.}\ \bibnamefont {Miller}},\ }\bibfield  {title} {\bibinfo {title} {{Light front nuclear physics: Mean field theory for finite nuclei}},\ }\href {https://doi.org/10.1103/PhysRevC.60.055211} {\bibfield  {journal} {\bibinfo  {journal} {Phys. Rev. C}\ }\textbf {\bibinfo {volume} {60}},\ \bibinfo {pages} {055211} (\bibinfo {year} {1999}{\natexlab{a}})},\ \Eprint {https://arxiv.org/abs/nucl-th/9906012} {arXiv:nucl-th/9906012} \BibitemShut {NoStop}%
\bibitem [{\citenamefont {Blunden}\ \emph {et~al.}(1999{\natexlab{b}})\citenamefont {Blunden}, \citenamefont {Burkardt},\ and\ \citenamefont {Miller}}]{Blunden:1999hy}%
  \BibitemOpen
  \bibfield  {author} {\bibinfo {author} {\bibfnamefont {P.~G.}\ \bibnamefont {Blunden}}, \bibinfo {author} {\bibfnamefont {M.}~\bibnamefont {Burkardt}},\ and\ \bibinfo {author} {\bibfnamefont {G.~A.}\ \bibnamefont {Miller}},\ }\bibfield  {title} {\bibinfo {title} {{Rotational invariance in nuclear light-front mean field theory}},\ }\href {https://doi.org/10.1103/PhysRevC.59.R2998} {\bibfield  {journal} {\bibinfo  {journal} {Phys. Rev. C}\ }\textbf {\bibinfo {volume} {59}},\ \bibinfo {pages} {2998} (\bibinfo {year} {1999}{\natexlab{b}})},\ \Eprint {https://arxiv.org/abs/nucl-th/9901063} {arXiv:nucl-th/9901063} \BibitemShut {NoStop}%
\bibitem [{\citenamefont {Vary}\ \emph {et~al.}(2010)\citenamefont {Vary}, \citenamefont {Honkanen}, \citenamefont {Li}, \citenamefont {Maris}, \citenamefont {Brodsky}, \citenamefont {Harindranath}, \citenamefont {de~Teramond}, \citenamefont {Sternberg}, \citenamefont {Ng},\ and\ \citenamefont {Yang}}]{Vary:2009gt}%
  \BibitemOpen
  \bibfield  {author} {\bibinfo {author} {\bibfnamefont {J.~P.}\ \bibnamefont {Vary}}, \bibinfo {author} {\bibfnamefont {H.}~\bibnamefont {Honkanen}}, \bibinfo {author} {\bibfnamefont {J.}~\bibnamefont {Li}}, \bibinfo {author} {\bibfnamefont {P.}~\bibnamefont {Maris}}, \bibinfo {author} {\bibfnamefont {S.~J.}\ \bibnamefont {Brodsky}}, \bibinfo {author} {\bibfnamefont {A.}~\bibnamefont {Harindranath}}, \bibinfo {author} {\bibfnamefont {G.~F.}\ \bibnamefont {de~Teramond}}, \bibinfo {author} {\bibfnamefont {P.}~\bibnamefont {Sternberg}}, \bibinfo {author} {\bibfnamefont {E.~G.}\ \bibnamefont {Ng}},\ and\ \bibinfo {author} {\bibfnamefont {C.}~\bibnamefont {Yang}},\ }\bibfield  {title} {\bibinfo {title} {{Hamiltonian light-front field theory in a basis function approach}},\ }\href {https://doi.org/10.1103/PhysRevC.81.035205} {\bibfield  {journal} {\bibinfo  {journal} {Phys. Rev. C}\ }\textbf {\bibinfo {volume} {81}},\ \bibinfo {pages} {035205} (\bibinfo {year} {2010})},\ \Eprint {https://arxiv.org/abs/0905.1411}
  {arXiv:0905.1411 [nucl-th]} \BibitemShut {NoStop}%
\bibitem [{\citenamefont {Brodsky}\ and\ \citenamefont {de~Teramond}(2006)}]{Brodsky:2006uqa}%
  \BibitemOpen
  \bibfield  {author} {\bibinfo {author} {\bibfnamefont {S.~J.}\ \bibnamefont {Brodsky}}\ and\ \bibinfo {author} {\bibfnamefont {G.~F.}\ \bibnamefont {de~Teramond}},\ }\bibfield  {title} {\bibinfo {title} {{Hadronic spectra and light-front wavefunctions in holographic QCD}},\ }\href {https://doi.org/10.1103/PhysRevLett.96.201601} {\bibfield  {journal} {\bibinfo  {journal} {Phys. Rev. Lett.}\ }\textbf {\bibinfo {volume} {96}},\ \bibinfo {pages} {201601} (\bibinfo {year} {2006})},\ \Eprint {https://arxiv.org/abs/hep-ph/0602252} {arXiv:hep-ph/0602252} \BibitemShut {NoStop}%
\bibitem [{\citenamefont {Brodsky}\ \emph {et~al.}(2015)\citenamefont {Brodsky}, \citenamefont {de~Teramond}, \citenamefont {Dosch},\ and\ \citenamefont {Erlich}}]{Brodsky:2014yha}%
  \BibitemOpen
  \bibfield  {author} {\bibinfo {author} {\bibfnamefont {S.~J.}\ \bibnamefont {Brodsky}}, \bibinfo {author} {\bibfnamefont {G.~F.}\ \bibnamefont {de~Teramond}}, \bibinfo {author} {\bibfnamefont {H.~G.}\ \bibnamefont {Dosch}},\ and\ \bibinfo {author} {\bibfnamefont {J.}~\bibnamefont {Erlich}},\ }\bibfield  {title} {\bibinfo {title} {{Light-Front Holographic QCD and Emerging Confinement}},\ }\href {https://doi.org/10.1016/j.physrep.2015.05.001} {\bibfield  {journal} {\bibinfo  {journal} {Phys. Rept.}\ }\textbf {\bibinfo {volume} {584}},\ \bibinfo {pages} {1} (\bibinfo {year} {2015})},\ \Eprint {https://arxiv.org/abs/1407.8131} {arXiv:1407.8131 [hep-ph]} \BibitemShut {NoStop}%
\bibitem [{\citenamefont {Freese}\ \emph {et~al.}(2026)\citenamefont {Freese}, \citenamefont {Lashbrook},\ and\ \citenamefont {Miller}}]{Freese:2026ulx}%
  \BibitemOpen
  \bibfield  {author} {\bibinfo {author} {\bibfnamefont {A.~J.}\ \bibnamefont {Freese}}, \bibinfo {author} {\bibfnamefont {A.~L.}\ \bibnamefont {Lashbrook}},\ and\ \bibinfo {author} {\bibfnamefont {G.~A.}\ \bibnamefont {Miller}},\ }\href@noop {} {\bibinfo {title} {{Three-dimensional, boost-invariant formalism for systems of relativistically moving constituents}}} (\bibinfo {year} {2026}),\ \Eprint {https://arxiv.org/abs/2607.12869} {arXiv:2607.12869 [nucl-th]} \BibitemShut {NoStop}%
\bibitem [{\citenamefont {Li}\ \emph {et~al.}(2016)\citenamefont {Li}, \citenamefont {Maris}, \citenamefont {Zhao},\ and\ \citenamefont {Vary}}]{Li:2015zda}%
  \BibitemOpen
  \bibfield  {author} {\bibinfo {author} {\bibfnamefont {Y.}~\bibnamefont {Li}}, \bibinfo {author} {\bibfnamefont {P.}~\bibnamefont {Maris}}, \bibinfo {author} {\bibfnamefont {X.}~\bibnamefont {Zhao}},\ and\ \bibinfo {author} {\bibfnamefont {J.~P.}\ \bibnamefont {Vary}},\ }\bibfield  {title} {\bibinfo {title} {{Heavy Quarkonium in a Holographic Basis}},\ }\href {https://doi.org/10.1016/j.physletb.2016.04.065} {\bibfield  {journal} {\bibinfo  {journal} {Phys. Lett. B}\ }\textbf {\bibinfo {volume} {758}},\ \bibinfo {pages} {118} (\bibinfo {year} {2016})},\ \Eprint {https://arxiv.org/abs/1509.07212} {arXiv:1509.07212 [hep-ph]} \BibitemShut {NoStop}%
\bibitem [{\citenamefont {Mayer}(1949)}]{Mayer1949}%
  \BibitemOpen
  \bibfield  {author} {\bibinfo {author} {\bibfnamefont {M.~G.}\ \bibnamefont {Mayer}},\ }\bibfield  {title} {\bibinfo {title} {On closed shells in nuclei. ii},\ }\href {https://doi.org/10.1103/PhysRev.75.1969} {\bibfield  {journal} {\bibinfo  {journal} {Physical Review}\ }\textbf {\bibinfo {volume} {75}},\ \bibinfo {pages} {1969} (\bibinfo {year} {1949})}\BibitemShut {NoStop}%
\bibitem [{\citenamefont {Haxel}\ \emph {et~al.}(1949)\citenamefont {Haxel}, \citenamefont {Jensen},\ and\ \citenamefont {Suess}}]{Haxel1949}%
  \BibitemOpen
  \bibfield  {author} {\bibinfo {author} {\bibfnamefont {O.}~\bibnamefont {Haxel}}, \bibinfo {author} {\bibfnamefont {J.~H.~D.}\ \bibnamefont {Jensen}},\ and\ \bibinfo {author} {\bibfnamefont {H.~E.}\ \bibnamefont {Suess}},\ }\bibfield  {title} {\bibinfo {title} {On the ``magic numbers'' in nuclear structure},\ }\href {https://doi.org/10.1103/PhysRev.75.1766.2} {\bibfield  {journal} {\bibinfo  {journal} {Physical Review}\ }\textbf {\bibinfo {volume} {75}},\ \bibinfo {pages} {1766} (\bibinfo {year} {1949})}\BibitemShut {NoStop}%
\bibitem [{\citenamefont {Wong}(1998)}]{Wong1998}%
  \BibitemOpen
  \bibfield  {author} {\bibinfo {author} {\bibfnamefont {S.~S.~M.}\ \bibnamefont {Wong}},\ }\href {https://doi.org/10.1002/9783527617906} {\emph {\bibinfo {title} {Introductory Nuclear Physics}}},\ \bibinfo {edition} {2nd}\ ed.\ (\bibinfo  {publisher} {John Wiley \& Sons},\ \bibinfo {year} {1998})\BibitemShut {NoStop}%
\bibitem [{\citenamefont {Miller}(2009{\natexlab{a}})}]{Miller:2009sg}%
  \BibitemOpen
  \bibfield  {author} {\bibinfo {author} {\bibfnamefont {G.~A.}\ \bibnamefont {Miller}},\ }\bibfield  {title} {\bibinfo {title} {{Electromagnetic Form Factors and Charge Densities From Hadrons to Nuclei}},\ }\href {https://doi.org/10.1103/PhysRevC.80.045210} {\bibfield  {journal} {\bibinfo  {journal} {Phys. Rev. C}\ }\textbf {\bibinfo {volume} {80}},\ \bibinfo {pages} {045210} (\bibinfo {year} {2009}{\natexlab{a}})},\ \Eprint {https://arxiv.org/abs/0908.1535} {arXiv:0908.1535 [nucl-th]} \BibitemShut {NoStop}%
\bibitem [{\citenamefont {Freese}\ and\ \citenamefont {Miller}(2023)}]{Freese:2023jcp}%
  \BibitemOpen
  \bibfield  {author} {\bibinfo {author} {\bibfnamefont {A.}~\bibnamefont {Freese}}\ and\ \bibinfo {author} {\bibfnamefont {G.~A.}\ \bibnamefont {Miller}},\ }\bibfield  {title} {\bibinfo {title} {{Light front synchronization and rest frame densities of the proton: Electromagnetic densities}},\ }\href {https://doi.org/10.1103/PhysRevD.107.074036} {\bibfield  {journal} {\bibinfo  {journal} {Phys. Rev. D}\ }\textbf {\bibinfo {volume} {107}},\ \bibinfo {pages} {074036} (\bibinfo {year} {2023})},\ \Eprint {https://arxiv.org/abs/2302.09171} {arXiv:2302.09171 [hep-ph]} \BibitemShut {NoStop}%
\bibitem [{\citenamefont {Miller}\ and\ \citenamefont {Brodsky}(2020)}]{Miller:2019ysh}%
  \BibitemOpen
  \bibfield  {author} {\bibinfo {author} {\bibfnamefont {G.~A.}\ \bibnamefont {Miller}}\ and\ \bibinfo {author} {\bibfnamefont {S.~J.}\ \bibnamefont {Brodsky}},\ }\bibfield  {title} {\bibinfo {title} {{Frame-independent spatial coordinate $\tilde{z}$: Implications for light-front wave functions, deep inelastic scattering, light-front holography, and lattice QCD calculations}},\ }\href {https://doi.org/10.1103/PhysRevC.102.022201} {\bibfield  {journal} {\bibinfo  {journal} {Phys. Rev. C}\ }\textbf {\bibinfo {volume} {102}},\ \bibinfo {pages} {022201} (\bibinfo {year} {2020})},\ \Eprint {https://arxiv.org/abs/1912.08911} {arXiv:1912.08911 [hep-ph]} \BibitemShut {NoStop}%
\bibitem [{\citenamefont {Freese}(2025)}]{AdamLecture}%
  \BibitemOpen
  \bibfield  {author} {\bibinfo {author} {\bibfnamefont {A.}~\bibnamefont {Freese}},\ }\href {https://www.youtube.com/watch?v=6iYPcNVQ6fE} {\bibinfo {title} {Introduction to light-front physics, lecture 4: Harmonic oscillator example}},\ \bibinfo {howpublished} {Lecture delivered to the ILCAC Seminar Series; video recording on YouTube} (\bibinfo {year} {2025}),\ \bibinfo {note} {published October 20, 2025. Relevant discussion beginning at 25:01}\BibitemShut {NoStop}%
\bibitem [{\citenamefont {Frosch}\ \emph {et~al.}(1968)\citenamefont {Frosch}, \citenamefont {Hofstadter}, \citenamefont {McCarthy}, \citenamefont {Noldeke}, \citenamefont {van Oostrum}, \citenamefont {Yearian}, \citenamefont {Clark}, \citenamefont {Herman},\ and\ \citenamefont {Ravenhall}}]{Frosch:1968zz}%
  \BibitemOpen
  \bibfield  {author} {\bibinfo {author} {\bibfnamefont {R.~F.}\ \bibnamefont {Frosch}}, \bibinfo {author} {\bibfnamefont {R.}~\bibnamefont {Hofstadter}}, \bibinfo {author} {\bibfnamefont {J.~S.}\ \bibnamefont {McCarthy}}, \bibinfo {author} {\bibfnamefont {G.~K.}\ \bibnamefont {Noldeke}}, \bibinfo {author} {\bibfnamefont {K.~J.}\ \bibnamefont {van Oostrum}}, \bibinfo {author} {\bibfnamefont {M.~R.}\ \bibnamefont {Yearian}}, \bibinfo {author} {\bibfnamefont {B.~C.}\ \bibnamefont {Clark}}, \bibinfo {author} {\bibfnamefont {R.}~\bibnamefont {Herman}},\ and\ \bibinfo {author} {\bibfnamefont {D.~G.}\ \bibnamefont {Ravenhall}},\ }\bibfield  {title} {\bibinfo {title} {{Electron Scattering Studies of Calcium and Titanium Isotopes}},\ }\href {https://doi.org/10.1103/PhysRev.174.1380} {\bibfield  {journal} {\bibinfo  {journal} {Phys. Rev.}\ }\textbf {\bibinfo {volume} {174}},\ \bibinfo {pages} {1380} (\bibinfo {year} {1968})}\BibitemShut {NoStop}%
\bibitem [{\citenamefont {Miller}(2009{\natexlab{b}})}]{Miller_2009}%
  \BibitemOpen
  \bibfield  {author} {\bibinfo {author} {\bibfnamefont {G.~A.}\ \bibnamefont {Miller}},\ }\bibfield  {title} {\bibinfo {title} {Electromagnetic form factors and charge densities from hadrons to nuclei},\ }\bibfield  {journal} {\bibinfo  {journal} {Physical Review C}\ }\textbf {\bibinfo {volume} {80}},\ \href {https://doi.org/10.1103/physrevc.80.045210} {10.1103/physrevc.80.045210} (\bibinfo {year} {2009}{\natexlab{b}})\BibitemShut {NoStop}%
\bibitem [{\citenamefont {Angeli}\ and\ \citenamefont {Marinova}(2013)}]{AngeliMarinova2013}%
  \BibitemOpen
  \bibfield  {author} {\bibinfo {author} {\bibfnamefont {I.}~\bibnamefont {Angeli}}\ and\ \bibinfo {author} {\bibfnamefont {K.~P.}\ \bibnamefont {Marinova}},\ }\bibfield  {title} {\bibinfo {title} {Table of experimental nuclear ground state charge radii: An update},\ }\href {https://doi.org/10.1016/j.adt.2011.12.006} {\bibfield  {journal} {\bibinfo  {journal} {Atomic Data and Nuclear Data Tables}\ }\textbf {\bibinfo {volume} {99}},\ \bibinfo {pages} {69} (\bibinfo {year} {2013})}\BibitemShut {NoStop}%
\end{thebibliography}%
\end{document}